\documentclass[9pt,conference]{IEEEtran}
\IEEEoverridecommandlockouts
\usepackage{cite}
\usepackage[scaled]{beramono}
\usepackage[T1]{fontenc}
\usepackage[linesnumbered,ruled,vlined]{algorithm2e}
\usepackage{amsmath,amssymb,amsfonts}
\usepackage{algorithmic}
\usepackage{bbding}
\usepackage{graphicx}
\usepackage{textcomp}
\usepackage{xcolor}
\usepackage[T1]{fontenc}
\usepackage{multicol}
\usepackage{tikz}
\usepackage{subcaption}
\usepackage{tcolorbox}
\tcbuselibrary{skins}
\usepackage{booktabs}
\usepackage{url}
\def\BibTeX{{\rm B\kern-.05em{\sc i\kern-.025em b}\kern-.08em
    T\kern-.1667em\lower.7ex\hbox{E}\kern-.125emX}}
\begin{document}

\newcommand{\project}{TokenSim\xspace}

\newtcolorbox{findingbox}[1][Finding]{%
    colback=gray!10, % 背景颜色为浅灰色
    colframe=black, % 边框颜色为灰色
    top=2mm, % 框顶部的间距
    bottom=2mm, % 框底部的间距
    left=2mm, % 框左侧的间距
    right=2mm, % 框右侧的间距
    arc=0mm, % 不使用圆角
    % enhanced, % 使用增强样式
    boxrule=0.5pt, % 设置边框粗细为0.5pt
    % coltitle=black, % 标题颜色
    % fonttitle=\bfseries, % 标题字体样式
    % title=#1, % 标题内容
    % attach boxed title to top left={yshift=-2mm,xshift=2mm}, % 标题位置
    % boxed title style={size=small,colframe=gray!50,colback=gray!20} % 标题框样式
}

\title{\project: Enabling Hardware and Software Exploration for Large Language Model Inference Systems
% \vspace{-1.5cm}
}
% \title{Conference Paper Title*\\
% {\footnotesize \textsuperscript{*}Note: Sub-titles are not captured for https://ieeexplore.ieee.org  and

\author{
\IEEEauthorblockN{1\textsuperscript{st} Feiyang Wu}
\IEEEauthorblockA{\textit{Beihang University} \\
Beijing, China \\
21373411@buaa.edu.cn}

\and

\IEEEauthorblockN{2\textsuperscript{nd} Zhuohang Bian}
\IEEEauthorblockA{\textit{Beihang University} \\
Beijing, China \\
22373017@buaa.edu.cn}

\and

\IEEEauthorblockN{3\textsuperscript{nd} Guoyang Duan}
\IEEEauthorblockA{\textit{Peking University} \\
Beijing, China \\
2200011004@stu.pku.edu.cn}

\and

\IEEEauthorblockN{4\textsuperscript{rd} Tianle Xu}
\IEEEauthorblockA{\textit{Peking University} \\
Beijing, China \\
2200011072@stu.pku.edu.cn}

\and

\IEEEauthorblockN{5\textsuperscript{th} Junchi Wu}
\IEEEauthorblockA{\textit{Peking University} \\
Beijing, China \\
1281236805@qq.com}

\and

\IEEEauthorblockN{6\textsuperscript{th} Teng Ma}
\IEEEauthorblockA{\textit{Renmin University of China \& Alibaba Group} \\
Beijing, China \\
sima.mt@alibaba-inc.com}

\and

\IEEEauthorblockN{7\textsuperscript{th} Yongqiang Yao}
\IEEEauthorblockA{\textit{Sensetime} \\
Shanghai, China \\
soundbupt@gmail.com}

\and

\IEEEauthorblockN{8\textsuperscript{th} Ruihao Gong}
\IEEEauthorblockA{\textit{Sensetime} \\
Beijing, China \\
gongruihao@sensetime.com}

\and

\IEEEauthorblockN{9\textsuperscript{th} Youwei Zhuo}
\IEEEauthorblockA{\textit{Peking University} \\
Beijing, China \\
youwei@pku.edu.cn}
}

\maketitle

\begin{abstract}
The increasing demand for large language model (LLM) serving has necessitated significant advancements in the optimization and profiling of LLM inference systems. As these models become integral to a wide range of applications, the need for efficient and scalable serving solutions has grown exponentially. This work introduces \project, a comprehensive hardware and software exploration system designed specifically for LLM inference. \project\ is characterized by its support for extensible system optimizations including scheduling and memory management. We validate the results with systems running with real-world datasets, achieving an error rate of less than $1\%$. Furthermore, \project\ facilitates various insightful explorations into the performance and optimization of LLM serving systems. 
The code is available at \url{https://github.com/pku-lemonade/TokenSim}.
\end{abstract}

% \begin{IEEEkeywords}
% component, formatting, style, styling, insert.
% \end{IEEEkeywords}

\section{Introduction}

Large Language Models (LLMs)\cite{llm}, such as ChatGPT\cite{chatgpt} and Gemini\cite{google_gemini}, have demonstrated impressive capabilities in understanding and generating human-like content, thus revolutionizing applications in chatbots\cite{google_bard, chatgpt} and programming assistants\cite{github_copilot, evalllmcode}. As LLMs gain popularity, the growing computation and memory demand of LLM serving becomes increasingly challenging.

To address the demands, researchers have introduced a variety of hardware and software optimizations.
On the hardware side, new accelerators emerge with various peak floating point performance and memory bandwidth. For example, near-data-processing accelerators like \cite{hynix-aim,upmem} leverages novel memory technology and achieves higher bandwidth than conventional memory systems, which makes them favorable for memory-intensive operators. 
On the inference system side, 
the optimizations primarily focus on two key areas: request scheduling and memory management.
Request scheduling optimizations aim to improve the utilization of the accelerator by batching requests of different lengths, while memory management optimizations tackle with memory footprint and memory access efficiency.
For example, vLLM~\cite{vllm} adopts optimizations in both categories, which lead to orders of magnitude of latency reduction. Therefore, these system optimizations have become the norm in real-world LLM serving systems. 
The hardware and software innovations also open up new opportunities in cross-stack optimizations. For example, when managing a cluster of novel hardware accelerators, it is intuitive to implement heterogeneity-aware scheduling policies in the inference system. 

% For example, Orca~\cite{orca} proposes continuous batching to handle concurrent requests.
% By incorporating new requests in a running batch, the dynamic batching can achieve over ten times higher throughput.
% As requests arrive and leave, the memory footprint of the model changes over time.
% For example, vLLM~\cite{vllm} leverages a paged attention technique that manages KV cache in non-contiguous memory blocks. When the request arrives, it allocates a new block instead of reserving a large chunk of memory to accommodate the maximum request length. 
% vLLM significantly improves memory efficiency by reducing memory allocation and fragmentation.

Due to the diversity and complexity of existing optimization methods, it is important for LLM practitioners to understand and predict the performance and resource demand of LLM inference systems. 
Currently, several simulators\cite{genz,llmcompass} support the performance modeling of different hardware. 
Unfortunately, they restrict the input to single request or single batch, reporting only two numbers for a test case: latency and memory usage. These metrics are not enough for both the developers and users of LLM inference. 
In real-world LLM inference, a system typically handles hundreds to thousands of concurrent requests from different users.
The user satisfaction is closely related with tail latency and it is important to obtain the distribution of the latency. 
As new requests enter and leave the system, memory usage also changes. 

% Additionally, testing and profiling new features based on complex and code-intensive LLM inference systems (such as vLLM) often requires significant development time. 

In this work, we address the lack of dynamic support by 
providing a highly modular and extensible simulator \project. As shown in table \ref{table:comparison}, \project has two key features: 
(i) dynamic LLM request input support sampled from real datasets; (ii) user-defined scheduling and memory management at operator-level granularity. With the introduction of these new features, \project can simulate QoS measures and generate detailed performance results, including the latency distribution and memory usage over time. 

% With these new features, we present several cases studies to enable hardware and software explorations. We first demonstrate that existing simulators produce extremely inaccurate performance numbers when evaluating dynamic LLM workloads due to lack of batching support in section \ref{cont_batching}.
% Then we conduct experiments with other software optimizations, including restricting GPU memory utilization ratio (Section \ref{ratio}), exploring prefill/decode worker ratio (Section \ref{dis_1} and the ideal  worker hardware type of decode workers in  disaggregated setting (Section \ref{dis_2}).
% With the help of \project memory management design, we can also test the performance implication of leveraging a memory pool for long conversations (Section \ref{mem}). 
% These experiments emphasizing the importance of system-level simulation. 
% Finally, we explore the impact of FLOPS, memory bandwidth, and memory capacity in a real LLM serving system(Section \ref{arch}), offering insights on the different hardware resources demand for prefill and decode workers. 

% GPT optimized ver.

We present several case studies to facilitate hardware and software exploration. First, we demonstrate that existing simulators yield highly inaccurate performance metrics for dynamic LLM workloads due to their lack of batching support, as discussed in Section \ref{cont_batching}.
Subsequently, we conduct experiments involving various software optimizations. These include limiting the GPU memory utilization ratio (Section \ref{ratio}), examining the worker ratio (Sections \ref{dis_1}), and identifying the optimal hardware type for decode workers in disaggregated settings (Section \ref{dis_2}).
Leveraging \project's memory management design, we also assess the performance implications of utilizing a memory pool for long conversations (Section \ref{mem}). These experiments underscore the significance of system-level simulation.
Finally, we investigate the impact of FLOPS, memory bandwidth, and memory capacity in a real LLM serving system (Section \ref{arch}), providing insights into the different hardware resource demands between prefill and decode workers.

\section{Background}

\subsection{LLM Inference}

% A large language model (LLM)\cite{llm} processes a sentence by encoding it into a sequence of tokens and predicting the subsequent tokens. These tokens are embedded into a latent vector, which is then processed through multiple transformer decoder blocks\cite{attention}, including an attention layer and a multi-layer perceptron (MLP). Within the attention layer, the latent vector is projected into three linear transformations, producing three activations: query (Q), key (K), and value (V). These activations are utilized to update the latent vector by incorporating attention to other tokens. 

% During inference, LLM generates one new token per iteration through this process. 
% By repeating the iteration until reaches a maximum length or encounters a stop token, the model constructs the entire sentence.

Typically, large language models (LLMs)\cite{llm} are composed of several transformer\cite{attention} decoder blocks, each containing a self-attention and a multi-layer perceptron (MLP). 
During the generation process, the LLM samples and generates new tokens one by one, with each token depending on all preceding tokens. 
Due to this sequential dependency, the key and value vectors generated by preceding tokens in self-attention are often cached for generating subsequent tokens, referred to as the key-value (KV) cache\cite{kvcache}.

\begin{table}[tbp]
\caption{Comparison of LLM Simulation Methods.}
\vspace{-2mm}
\begin{center}
% \resizebox{0.5\textwidth}{10mm}{
\setlength{\tabcolsep}{1mm}{
\begin{tabular}{l|cccccc}
\hline
\textbf{Methods} & \textbf{Modular} & \textbf{Scheduler} & \multicolumn{1}{l}{\textbf{Mem Manager}} & \textbf{Portable} & \textbf{Dataset} \\ \hline
Roofline          & \XSolidBrush & \XSolidBrush & \XSolidBrush & \Checkmark   & \XSolidBrush \\
GenZ              & \XSolidBrush & \XSolidBrush & \XSolidBrush & \XSolidBrush & \XSolidBrush \\
LLMCompass        & \XSolidBrush & \XSolidBrush & \XSolidBrush & \XSolidBrush & \XSolidBrush \\
\textbf{\project} & \Checkmark   & \Checkmark   & \Checkmark   & \Checkmark   & \Checkmark   \\ \hline
\end{tabular}}
% }
\end{center}
\vspace{-4mm}
\label{table:comparison}
\end{table}

% Since the query and key activations for previous tokens remain constant across iterations, they only need to be computed once during the initial iteration. Their values can then be stored as key-value (KV) cache for subsequent use. This initial computation of all query and key activations is referred to as the prefill stage. 
% The LLM then uses the KV cache to predict the next token, storing its KV cache, and repeats this process for further generation. This iterative process for generating new tokens is known as the decode stage.
% The prefill stage, characterized by extensive matrix multiplications, is strongly compute-bound. Conversely, the decode stage requires frequent access to the KV cache in GPU memory during each iteration, making it memory-bound.

Depending on the cauculation method, the inference process is often divided into two stages: 
In the \textbf{prefill stage}, each request runs only one iteration to generate the KV cache for the prompt part. This stage involves parallel computation of each token in the prompt through matrix multiplication, making it highly \textbf{compute-bound}. 
In the \textbf{decode stage}, the LLM utilizes the previously generated KV cache to autoregressively generate subsequent tokens over several iterations. This stage involves matrix-vector multiplication, calculating only one new token at a time, thus making it highly \textbf{memory-bound}.

\vspace{-1mm}
\subsection{LLM Inference System Optimizations}

\noindent\textbf{Continuous batching}\cite{orca} is a technique employed in the inference of LLMs to enhance throughput and efficiency. 
% In traditional batching, requests are accumulated until a batch is full before processing, which can lead to delays, especially when requests arrive sporadically.
In traditional static batching, requests are accumulated and not processed until a batch is full, which often lead to delays.
% Continuous batching, on the other hand, dynamically adjusts the batch size based on the current workload, allowing for immediate processing of incoming requests without waiting for a batch to fill completely. 
Continuous batching, on the other hand, allows for immediate processing of incoming requests by dynamically adjusting the batch size between iterations.
% This approach is particularly beneficial in scenarios with variable request lengths, as it minimizes latency and maximizes resource utilization by ensuring that the computational resources are continuously engaged.
This approach achieves high GPU utilization through fine-grained batch adjustment, allowing requests to be processed immediately upon arrival, thereby significantly improved request latency.

% \noindent\textbf{PagedAttention}\cite{vllm} is an advanced memory management technique designed to optimize the handling of KV cache during LLM inference. As the context length increases, the memory required to store KV cache can become substantial, particularly on hardware with limited memory capacity. PagedAttention addresses this challenge by organizing the KV cache into pages, similar to the paging mechanism used in operating systems for virtual memory management. This approach effectively eliminates external fragmentation by ensuring that memory allocation and deallocation occur in fixed-size units named blocks, maintaining contiguous blocks of free memory, and minimizes internal fragmentation by allowing precise memory allocation that closely matches the actual needs of the KV cache. Consequently, PagedAttention optimizes memory utilization, enhancing the scalability and efficiency of LLM inference systems, particularly in environments with constrained memory resources.

\noindent\textbf{PagedAttention} \cite{vllm} is an advanced technique for managing KV cache during inference.
It mitigates memory fragmentation caused by varying request and output lengths by partitioning GPU memory into blocks and mapping logical to physical blocks, similar to the operating system memory management.
This approach enhances memory efficiency and significantly improves overall throughput.

\noindent\textbf{Disaggregated serving}\cite{splitwise,distserve} is a recent optimization that focus on the distinct characteristics of prefill and decode stages.
By disaggregating the stages across different devices, it allows for optimized resource allocation, 
% leveraging the compute-intensive nature of the prefill stage and the memory-intensive demands of the decode stage, 
thereby improving overall efficiency in LLM inference.
Based on the disaggregated architectures, in order to reuse KV cache between multi-round conversations, \textbf{CachedAttention}\cite{cachedattention} introduces a novel system integrating context caching based on disaggregated architecture.
CachedAttention maintains a caching system that leverages efficient memory and storage mediums to save KV caches for all requests.
\textbf{MemServe}\cite{memserve} implements a similar system with a memory pool managing distributed KV cache through APIs, while enhancing cache reuse through a global prompt tree-based locality-aware policy.
Both approaches improve the efficiency of serving multi-round conversations by optimizing cache management and reuse strategies.

\subsection{Simulating Frameworks}

To efficiently and quickly conduct targeting tests, some studies\cite{genz}\cite{llmcompass} have been proposed to provide simulations specifically for LLM inference tasks. 
\textbf{GenZ}\cite{genz} offers the simulation of single LLM inference iteration for different parallel methods and hardware parameters.%, providing detailed insights into the time consumption of each computational and communication step.
% This allows for targeted insights into different hardware parameters for prefill and decode stages of LLM inference. 
\textbf{LLMCompass}\cite{llmcompass} goes a step further by simulating single-layer computation of LLM inference based on different systolic arrays and buffer designs.%, offering hardware design space exploration for efficient LLM inference and providing optimal matrix computation designs.

% However, both approaches only implement static batching and cannot support complex schedule methods like continuous batching.
% Their memory management method is also very simple, with only calculations for KV cache sizes and cannot support PagedAttention.
% And both simulators runs on fixed input size without support of dataset reading and simulation for real-time serving.
% Thus, even though some insights have been proposed, they are far from practical application and struggle to support simulations of new optimizations based on existing frameworks. 
% If a new optimization based on vLLM\cite{vllm} requires simulation, these approaches are unlikely to provide the necessary simulation support due to the simplicity.

Recent advancements in the field have introduced support for sophisticated scheduling techniques, such as continuous batching, and advanced memory management strategies, like PagedAttention.
% Furthermore, the simulators use fixed input sizes without real-time serving simulation capabilities.
% As a result, they fall short of practical application and cannot effectively simulate new optimizations, such as those based on vLLM \cite{vllm}, due to their inherent simplicity.
Vidur~\cite{vidur} is a large-scale simulation framework that supports advanced scheduling methods and memory management, following the approach of vLLM~\cite{vllm}. However, it relies on random forest regression models to estimate calculation runtime, which may introduce additional errors. Meanwhile, LLMServingSim~\cite{llmservingsim} provides a hardware-software co-simulation infrastructure but suffers from significant performance limitations and lacks flexibility for customizing new models and hardware architectures.

Our \project employs a detailed transformer-oriented simulation model to address these limitations, achieving higher accuracy while maintaining faster simulation speeds. 
Moreover, \project supports novel techniques, such as simulating disaggregated prefill and decoding phases.\cite{distserve}
This approach enables comprehensive precision improvements without sacrificing efficiency.

\section{\project Design}

\begin{figure}[htbp]
    \vspace{1mm}
\centerline{\includegraphics[width=1\linewidth]{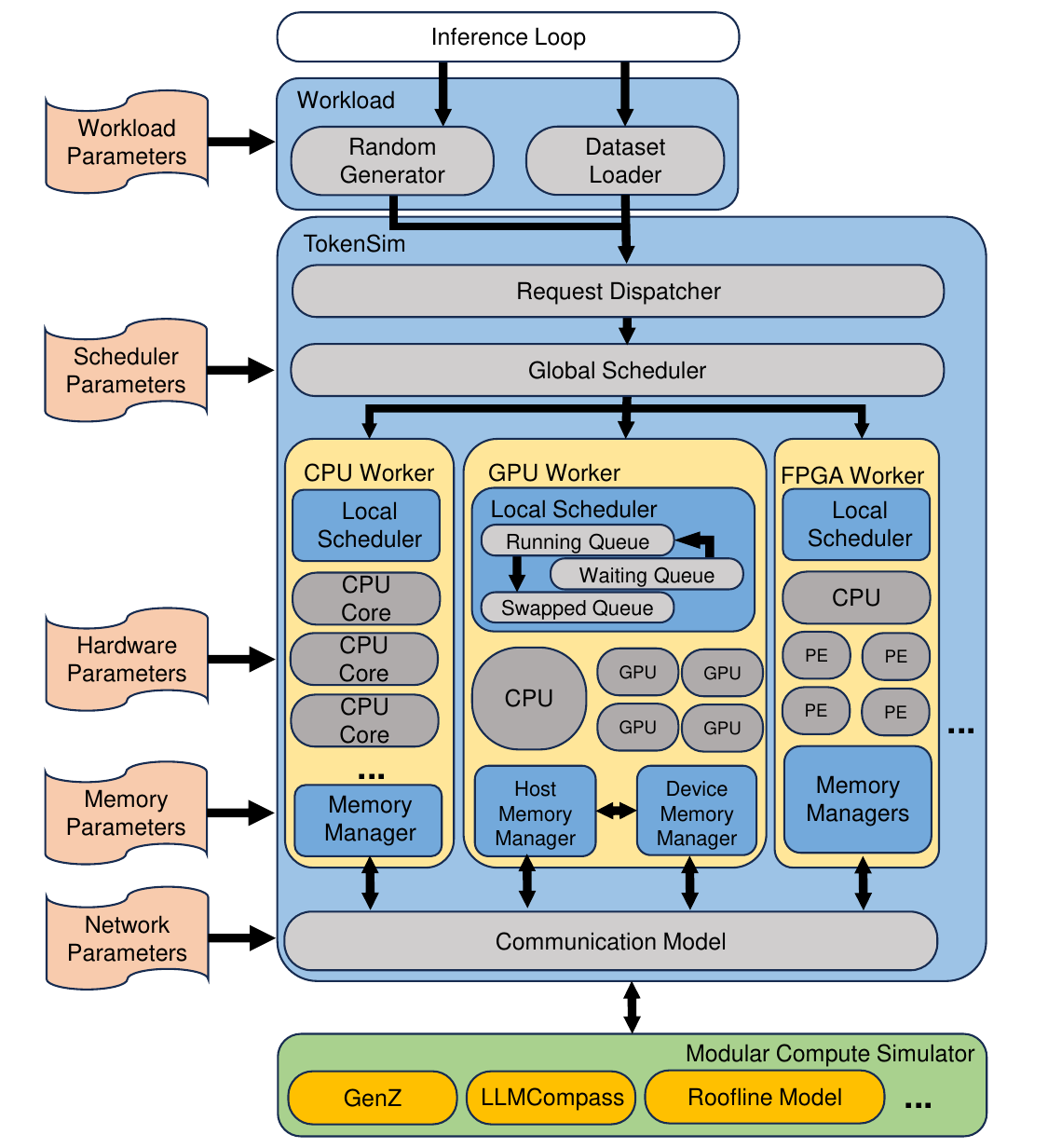}}
    \vspace{1mm}
    \caption{\project\ system overview.}
    \label{fig:overview}
    % \vspace{-1mm}
\end{figure}

In this section, we describe the internal implementation of \project and demonstrate how users can define their own scheduling and memory management policies in a modular and extensible manner. 

\project utilizes the SimPy\cite{simpy} library to conduct simulations, leveraging its event-driven architecture to model complex systems efficiently. SimPy is a lightweight, discrete-event simulation framework that allows for the modeling of active components, such as workers and devices, working parallel with time tracked within an simulated environment. SimPy empowers \project the ability to simulate processes in a highly efficient manner, supporting running even on personal computers without GPU.

Figure \ref{fig:overview} shows the architecture of \project. 
% The system operates within an inference loop that tracks simulated time across all workers. \project\ generates workloads from datasets and parameters that define the distribution and rate. Requests are dispatched at different time by a dispatcher to the global scheduler, which then allocates these requests to various workers based on user-defined scheduling settings.
The system operates within an inference loop tracking simulated time across all workers. \project\ generates workloads from datasets and parameters, with requests dispatched by a dispatcher to the global scheduler, which allocates them to workers based on user-defined settings.

% \project\ operates by running each worker simultaneously, with request management facilitated by each worker's local scheduler. These local schedulers employ various scheduling algorithms to schedule batches and iterations for request processing. Concurrently, the schedulers interface with memory managers that monitor memory utilization for each device. Once a batch is formed for an iteration, the relevant batch information is transmitted to a modular compute simulator, such as GenZ, to determine the iteration time. The system's architecture allows for the seamless integration of different compute simulators, thereby supporting a wide range of hardware configurations and simulation methods. For inter-device data movement, the communication model connects different memory managers for calculating and simulating transfer latency based on network parameters.
Each worker runs concurrently, managed by local schedulers employing various algorithms. These schedulers coordinate with memory managers that monitor device memory utilization. Once a batch is formed by scheduler for an iteration, relevant information is sent to a compute simulator, like GenZ, to determine iteration time. The architecture supports diverse compute simulators, enabling various hardware configurations and simulation methods. When inter-device data movement occurs, the communication model calculates transfer latency by connecting memory managers based on network parameters.

\begin{figure}[t]
\centering
\vspace{-2mm}
\centerline{\includegraphics[width=1\linewidth]{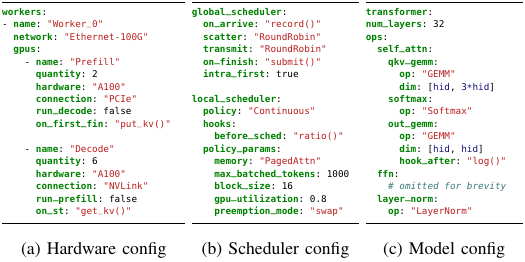}}
\vspace{1mm}
\caption{One example of \project\ configurations.}
\vspace{-4mm}
\label{fig:codes}
\end{figure}

% Scheduler.
% Memory Manager.
% Computation and Communication Modeling. 
% Simulator Input. 
% All requests generated by the workload generator are dispatched in the global scheduler to different workers. 

\subsection{\project Scheduler}

% Existing LLM inference serving systems employ sophisticated intra-worker (continous batching) and inter-worker (disaggregation) scheduling techniques. 
% In order to support both current and future scheduling innovations, \project implements a two-stage scheduler with one global scheduler and local schedulers for each worker.
% The global scheduler handles all requests when they first enter the system and assign them to a worker with a local scheduler. 
% After a request is assigned, the local scheduler decide if the request should continue to run on the local worker, or be transferred back to the global scheduler.

Existing LLM inference systems employ advanced intra-worker (continous batching) and inter-worker (disaggregation) scheduling techniques.
To support both current and future innovations, \project features a \textbf{two-stage scheduler} with a global scheduler and local schedulers for each worker.
The global scheduler manages incoming requests, assigning them to workers, while local schedulers decide between iterations whether requests should remain local or return to the global scheduler.

% \project allows user-defined functions for both global and local schedulers by modifying configuration files as illustrated in figure \ref{code:sched}. 
User-defined functions for both schedulers can be configured by modifying configuration file as shown in Figure \ref{code:sched}.
The scheduler function API provides all system information.
For example, the global scheduler can access the number of current workers, their hardware type, and concurrent requests. It can also be stateful, so that users can actively store the number of requests already dispatched to a worker in a given time window, and uses the record book for future load-aware scheduling.
For the local scheduler, the most important information needed in scheduling algorithm is the current task queue and memory utilization status. Fortunately, \project is able to simulate memory changes in real time, as will be described in the next section.

% To take full control of the scheduling granularity
% \project provide a user-defined \textbf{breakpoint} feature that allows programmers to set operator-level scheduler hooks in model configuration file in figure \ref{code:model}. By default, all requests have a hidden breakpoint at the end of an iteration, which means that the scheduler algorithm will be invoked after each time a request is advanced by one token. 
% With user-defined scheduling and breakpoints, it is easy to implement  popular inference optimizations. 
% For example, the disaggregation technique is essentially adding a breakpoint at the end of the first iteration for prefill devices and another breakpoint at the start of each iteration for decode devices, enabling KV cache transfer through two lines of code.

To take full control of scheduling granularity, \project introduces a \textbf{breakpoint} feature, enabling operator-level hooks in model configurations (Figure \ref{code:model}).
Default breakpoints invoke the scheduler after each token generation.
This setup simplifies implementing optimizations like disaggregation, which uses breakpoints for KV cache transfer with two lines of code, as shown in \ref{code:hard}.

% In \project, we can add an explicit breakpoint at the end of the model configuration file, and associate the breakpoint with: 1) a user-defined local scheduler that checks the current iteration count of the request and send all requests with one new token back to the global scheduler, 2) a user-defined global scheduler that receives requests from local schedulers and send request with one new token to one of the decode devices, as illustrated in figure \ref{code:python}.

\begin{figure}[t]
% \vspace{-2mm}
\centerline{\includegraphics[width=1\linewidth]{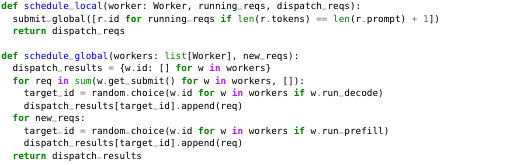}}
\vspace{1mm}
\caption{User-defined scheduler example of global scheduler and local scheduler, defining a disaggregated architecture.}
\vspace{-3mm}
\label{code:python}
\end{figure}

In \project, breakpoints can be explicitly added at the end of the model configuration file and linked to: 
1) a local scheduler that returns requests with new tokens to the global scheduler, and 
2) a global scheduler that dispatches these requests to decode devices, as shown in Figure \ref{code:python}.

% One of the primary objectives of \project\ is to facilitate developers in experimenting with and testing new designs. To this end, the system also offers a customizable scheduler besides the default vLLM scheduler. Through a provided template configuration and some simple Python codes, developers can create testable scheduling expressions without the need for managing complex hardware compatibilities.

% \IncMargin{1em}
% \begin{algorithm}

%     \SetKwInOut{Input}{Input}
%     \SetKwInOut{Output}{Output}

%     \Input{None}
%     \Output{A list of requests to run}

%     \BlankLine

%     \emph{Initialize} running \emph{as an empty list}\;
%     self.running.extend(self.waiting)\;
%     \emph{sort self.running by prompt length}\;

%     \While{\emph{self.running is not empty}}{
%         running.append(self.running.pop())\;
%         \If{sum(\emph{req.prompt\_len for req in running}) = \emph{self.max\_sum\_len}}{
%             \emph{break}\;
%         }
%     }

%     self.waiting $\leftarrow$ \emph{requests in self.running not in running}\;
%     self.running $\leftarrow$ running\;

%     \Return{running}\;

%     \caption{Long-Prompt-First Scheduling}
%     \label{algo:schedule}

% \end{algorithm}
% \DecMargin{1em}

\subsection{\project Memory Manager}

% Current techniques such as PagedAttention has introduced significant complexities in memory management. To effectively support memory managing innovations, \project have implemented various memory managers tailored to accommodate different types of workers: CPU worker, GPU worker, FPGA worker, etc. The goal of the memory managers is to monitor memory utilization at any granularity, by block, by token or by byte, supporting the user-defined scheduler behaviors.

Current techniques like PagedAttention have added significant complexity to memory management. To support these innovations, \project implements memory managers for various worker types, such as CPU, GPU, and FPGA, to monitor memory utilization at any granularity—by block, token, or byte—supporting user-defined scheduler behaviors.

Moreover, in a distributed environment, the memory of different workers are connected with network and inference systems tend to share them to create a large memory pool and transfer data from the memory of one node to another. For example, disaggregated architectures transfer KV cache between prefill and decode devices.

\project uses a communication component to model the data transfer overhead.   
This model takes cache location (host or device), data size and memory bandwidth as arguments and returns the time to transfer the data to another devices. 

This model is designed to support various types of memory transfer overlapping, leveraging SimPy's event-driven capabilities to facilitate both sequential and concurrent data loading and storage. By selecting appropriate data transfer sizes and overlap methods, simulations can be easily implemented.

For instance, in scenarios where data blocks are transferred from low-bandwidth to high-bandwidth storage, the default method initiates loading only after the receiver completes the previous task, executing each load and store operation sequentially. Furthermore, with more complex overlapping techniques, such as utilizing a preloading buffer, the communication model can achieve efficient data transfer by maintaining optimal buffer utilization. For an ideal transfer process that supports multiple data preloading, adjusting the model's buffer size to a larger value demonstrates the model's compatibility and simplicity.

\subsection{Validation Studies}

In this section, we validate the ouput of \project with performance results generated from real hardware with vLLM v0.6.2.

\begin{figure}[htbp]
    \vspace{-3mm}
    \centerline{\includegraphics[width=1.1\linewidth]{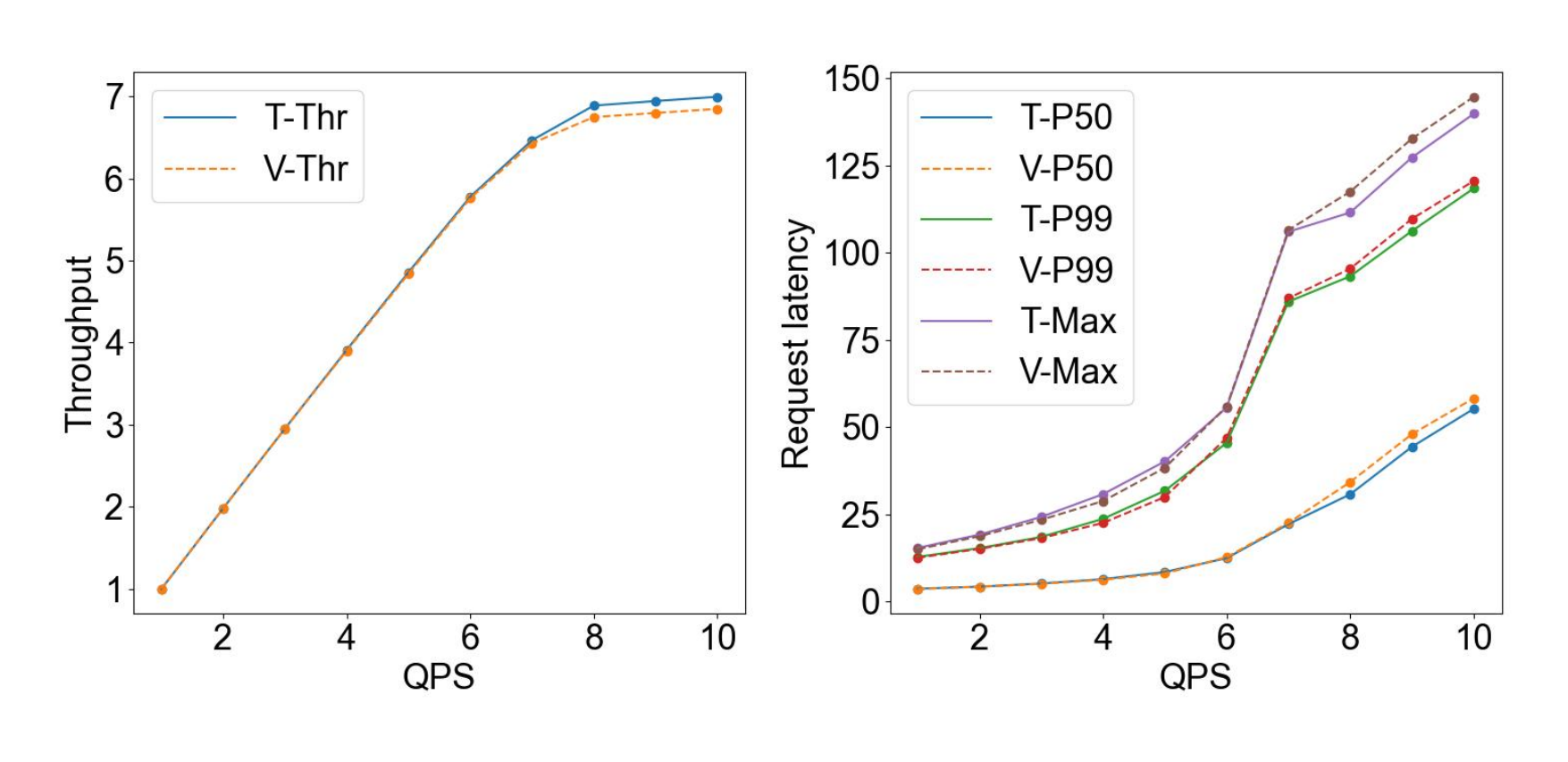}}
    \vspace{-1mm}
    \caption{vLLM throughput and latency validation. "V-" and "T-" stands for vLLM and \project respectively. "Thr" denotes throughput.}
    \vspace{-1mm}
    \label{fig:vllm1}
\end{figure}

\begin{figure}[htbp]
    \vspace{-3mm}
    \centerline{\includegraphics[width=\linewidth]{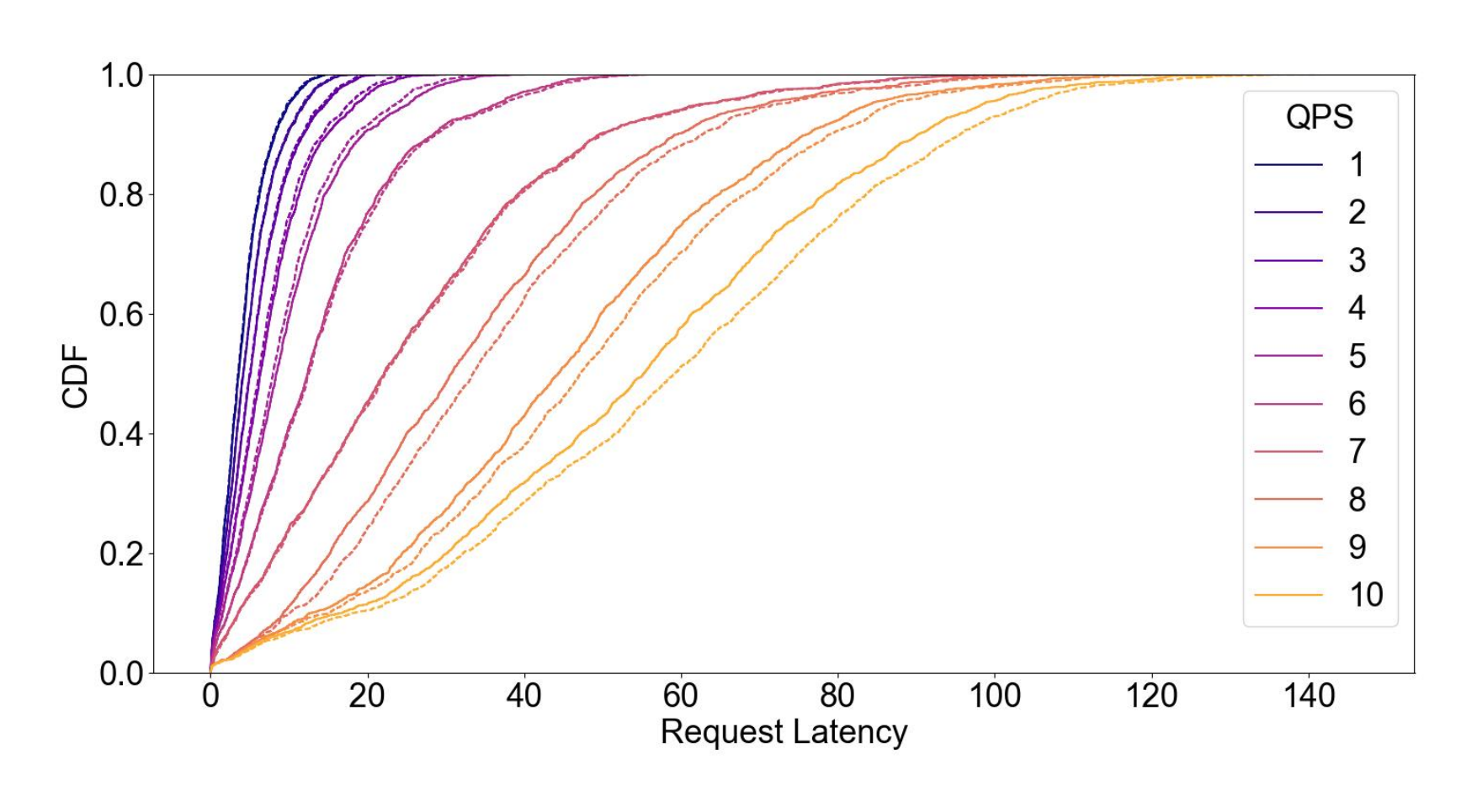}}
    % \vspace{-2mm}
    \caption{vLLM latency CDF aligns with \project at different QPS, dashed lines are vLLM and solid lines are \project.}
    \label{fig:vllm2}
    % \vspace{-2mm}
\end{figure}

% To evaluate the accuracy of \project, we performed a series of experiments serving LLaMA2-7B model on an NVIDIA A100 GPU with a dataset consisting of 2,000 requests sampled from ShareGPT dataset\cite{sharegpt}. By varying the request rates (queries per second, QPS), we compared the throughput and latency percentiles derived from \project's simulations with those obtained from vLLM's real-world results. The comparative results are presented in figure \ref{fig:vllm1}. The geometric mean error between the real and modeled values was calculated to be $0.109\%$ for throughput, and $0.6\%$, $0.254\%$, and $0.337\%$ for the P50, P99, and maximum request latency, respectively.

% To further validate the precision of the simulation, we recorded the latency of each request and plotted the cumulative distribution function (CDF) of the request latency distribution, as shown in figure \ref{fig:vllm2}. The close alignment between \project's simulation results and the observations from actual hardware demonstrates a high degree of accuracy.

To evaluate \project's accuracy, we conducted experiments with the LLaMA2-7B model\cite{llama, llama2} on an NVIDIA A100 GPU, using 2,000 requests from the ShareGPT dataset \cite{sharegpt}. By varying request rates (queries per second, QPS), we compared throughput and latency percentiles from \project's simulations with vLLM's real-world results, as shown in Figure \ref{fig:vllm1}. The geometric mean error is $0.109\%$ for throughput and $0.6\%$, $0.254\%$, and $0.337\%$ for P50, P99, and maximum request latency, respectively.

To further validate simulation precision, we recorded request latencies and plotted the cumulative distribution function (CDF) of the latency distribution, as illustrated in Figure \ref{fig:vllm2}. The close alignment between \project's simulations and actual hardware observations confirms a high degree of accuracy.

% \begin{figure}[htbp]
%     \centerline{\includegraphics[width=0.5\textwidth]{genz_compare.pdf}}
%     \caption{GenZ Compare result (tmp name)}
%     \label{fig:genz}
% \end{figure}

% \textbf{Validation Against Existing Frameworks.} To ensure that our simulation is consistent with existing studies, we compared the per-iteration time of \project\ with that of GenZ \cite{genz}. While \project\ is typically employed with varying request arrival times and diverse input lengths, GenZ is limited to static batching with single iteration simulations. Consequently, we recorded the single iteration time on \project\ using different batch sizes and input/output lengths, without executing the serving simulation. The results, presented in Figure \ref{fig:genz}, demonstrate that \project\ closely mirrors GenZ's simulation outcomes, particularly in the latency curves of both the prefill and decode stages, using the LLaMA2-7B model on a single A100 GPU.

\subsection{Comparison with State-of-the-Art Simulators}

We conducted a comprehensive comparison of \project against two state-of-the-art LLM inference simulators: Vidur \cite{vidur} and LLMServingSim \cite{llmservingsim}. Our evaluation focused on two key metrics: simulation accuracy and runtime efficiency.
Moreover, we test \project on its ability to simulate disaggregated prefill and decoding phases.

\subsubsection{Simulation Accuracy}

To evaluate the simulation accuracy of \project, we compared it with Vidur and LLMServingSim. The experimental setup is as follows: We conducted the experiments on an A100 GPU with 80 GB of memory. We first determined the optimal query-per-second (QPS) value by finding the point at which the LLM throughput reached 40 QPS, indicating the best performance state of the LLM. After obtaining this QPS value, we measured the total time elapsed from the submission of the first request to completion, varying the number of requests from 100 to 500.

Since the open-source version of LLMServingSim can only handle very short requests, we compared the results of all three simulators with the real-world scenario. We then compared \project and Vidur with the real-world scenario using fixed token lengths ranging from 128 to 2048.

As shown in Table~\ref{tab:perf_compare}, compared with the real results, \project demonstrated higher accuracy than both Vidur and LLMServingSim, representing a significant improvement over the state-of-the-art level.

We attribute our high simulation accuracy to our fine-grained memory simulation.
Through rigorous implementation and validation, we support block-granularity simulation, which provides a more detailed insight into the runtime status within Transformers.
By adopting this approach, we avoid coarse-grained approximations in MLP simulations, thereby more accurately reflecting real-world runtime scenarios.

\begin{table}[ht]
\centering
\caption{Percentage difference in latency between real hardware results and simulations from different simulators. The bars represent the percentage error in latency for 10 output tokens across different request counts, ranging from 100 to 500.}
\label{tab:perf_compare}
\begin{tabular}{l|ccccc}
\toprule
\textbf{Request num} & 100 & 200 & 300 & 400 & 500 \\
\hline \hline
% \textbf{Local} & 2.756 & 5.246 & 7.819 & 10.371 & 12.981 \\
% \textbf{Vidur} & 2.371\textbf{\scriptsize{(14.0\%)}} & 4.698\textbf{\scriptsize{(10.5\%)}} & 7.246\textbf{\scriptsize{(7.3\%)}} & 9.935\textbf{\scriptsize{(402\%)}} & 12.122\textbf{\scriptsize{(6.6\%)}} \\
% \textbf{\project} & 2.592\textbf{\scriptsize{(6.0\%)}} & 5.089\textbf{\scriptsize{(3.0\%)}} & 7.587\textbf{\scriptsize{(3.0\%)}} & 10.095\textbf{\scriptsize{(2.7\%)}} & 12.593\textbf{\scriptsize{(3.0\%)}} \\
% \textbf{LLMServingSim} & 2.556\textbf{\scriptsize{(7.3\%)}} & 5.056\textbf{\scriptsize{(3.6\%)}} & 7.557\textbf{\scriptsize{(3.4\%)}} & 10.056\textbf{\scriptsize{(3.0\%)}} & 12.556\textbf{\scriptsize{(3.3\%)}} \\
\textbf{Local} & 2.756 & 5.246 & 7.819 & 10.371 & 12.981 \\
\textbf{Vidur} & 2.371 & 4.698 & 7.246 & 9.935 & 12.122 \\
\textbf{\project} & 2.592 & 5.089 & 7.587 & 10.095 & 12.593 \\
\textbf{LLMServingSim} & 2.556 & 5.056 & 7.557 & 10.056 & 12.556 \\
\bottomrule
\end{tabular}
\end{table}

\subsubsection{Runtime Efficiency}

\begin{figure}[htbp]
    \vspace{-2mm}
    \centerline{\includegraphics[width=1\linewidth]{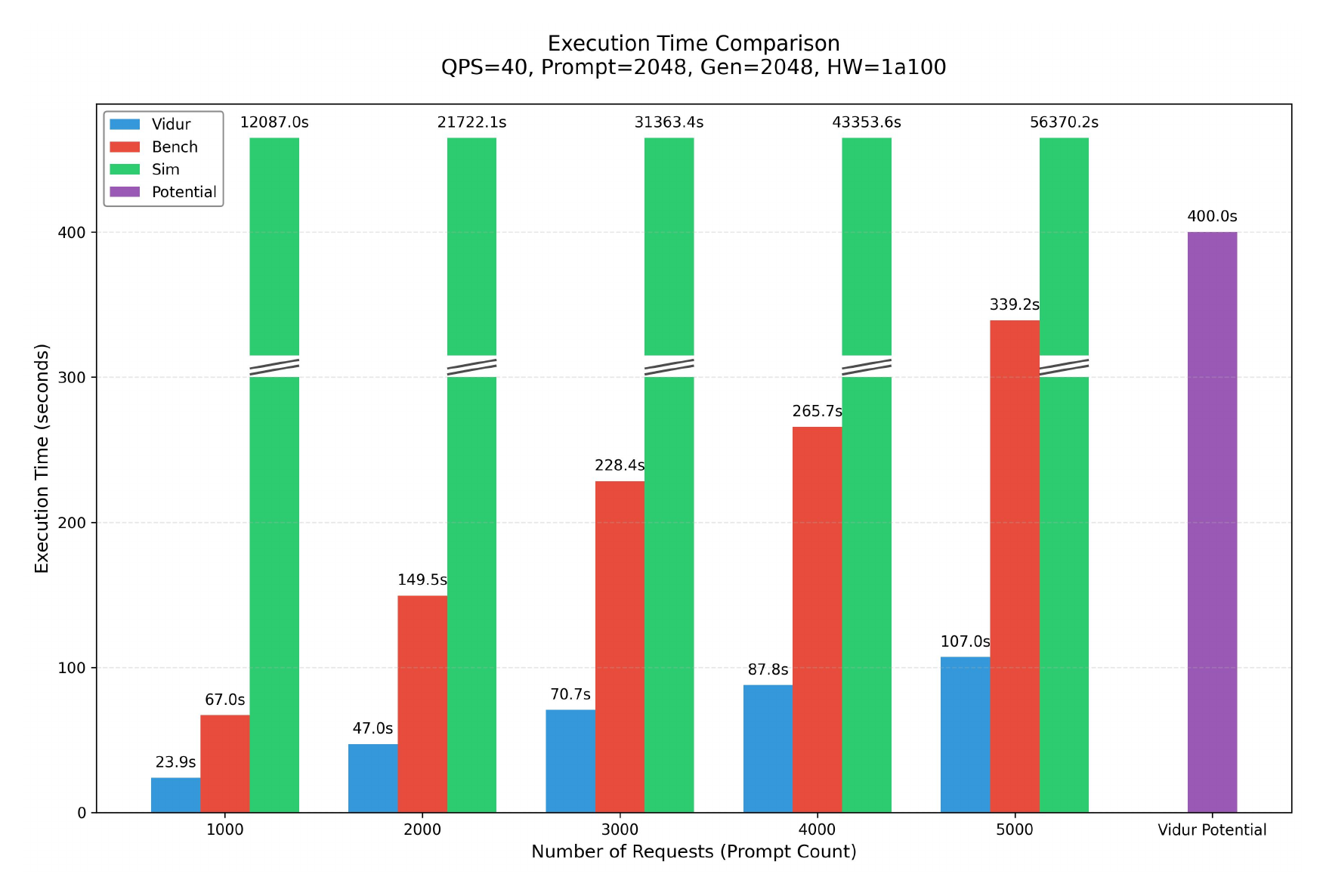}}
    \vspace{-1mm}
    \caption{Comparison of execution time between \project and Vidur. Note that Vidur requires significant pre-training time (approximately 400s) before each run, indicated by the shaded regions. Additionally, LLMServingSim is configured to handle only 10 tokens due to its inability to simulate long prompts, and it exhibits notably slow performance.}
    \label{fig:execution_time}
    % \vspace{-1mm}
\end{figure}

Regarding execution time, we compare our \project with Vidur. LLMServingSim is impressively slow, even slower than the real-time behavior. As can be seen in Figure~\ref{fig:execution_time}, although our \project appears to have longer runtimes, Vidur requires a significant amount of time---about 400 seconds for pre-training before each run.
Additionally, \project supports fine-grained memory operation simulations, offering more detailed insights. LLMServingSim is set to handle only 10 tokens due to its limitation in simulating long prompts, and it exhibits notably slow performance.

While \project may take slightly more time than Vidur, it avoids the pre-training procedure and is more lightweight. 
During the phase of verifying model runtime configurations, parameters may undergo substantial adjustments. The presence of numerous different configurations can render randomization ineffective. 
In terms of overall expected time, \project still holds an advantage, with its lightweight operation allowing for more flexible adaptation.

\subsubsection{Disaggregated Prefill and Decoding Phases}

We select DistServe~\cite{distserve} as our real-world baseline. 
To the best of our knowledge, \project is the first large language model simulator to support the disaggregation of prefill and decoding phases.

To validate this capability, we implement DistServe on two A100 GPUs.
We measure the actual communication bandwidth between the GPUs and use this data to configure \project for accurate simulation.
We then compare \project with DistServe using a set of requests ranging from 1000 to 10000, each with a fixed input token length of 64 tokens.
We selected a QPS value of 8 to minimize the runtime variations caused by different memory scheduling strategies, thereby focusing more effectively on the simulation performance of Prefilling and Decoding separation itself.

\begin{figure}[htbp]
    \vspace{-2mm}
    \centerline{\includegraphics[width=1\linewidth]{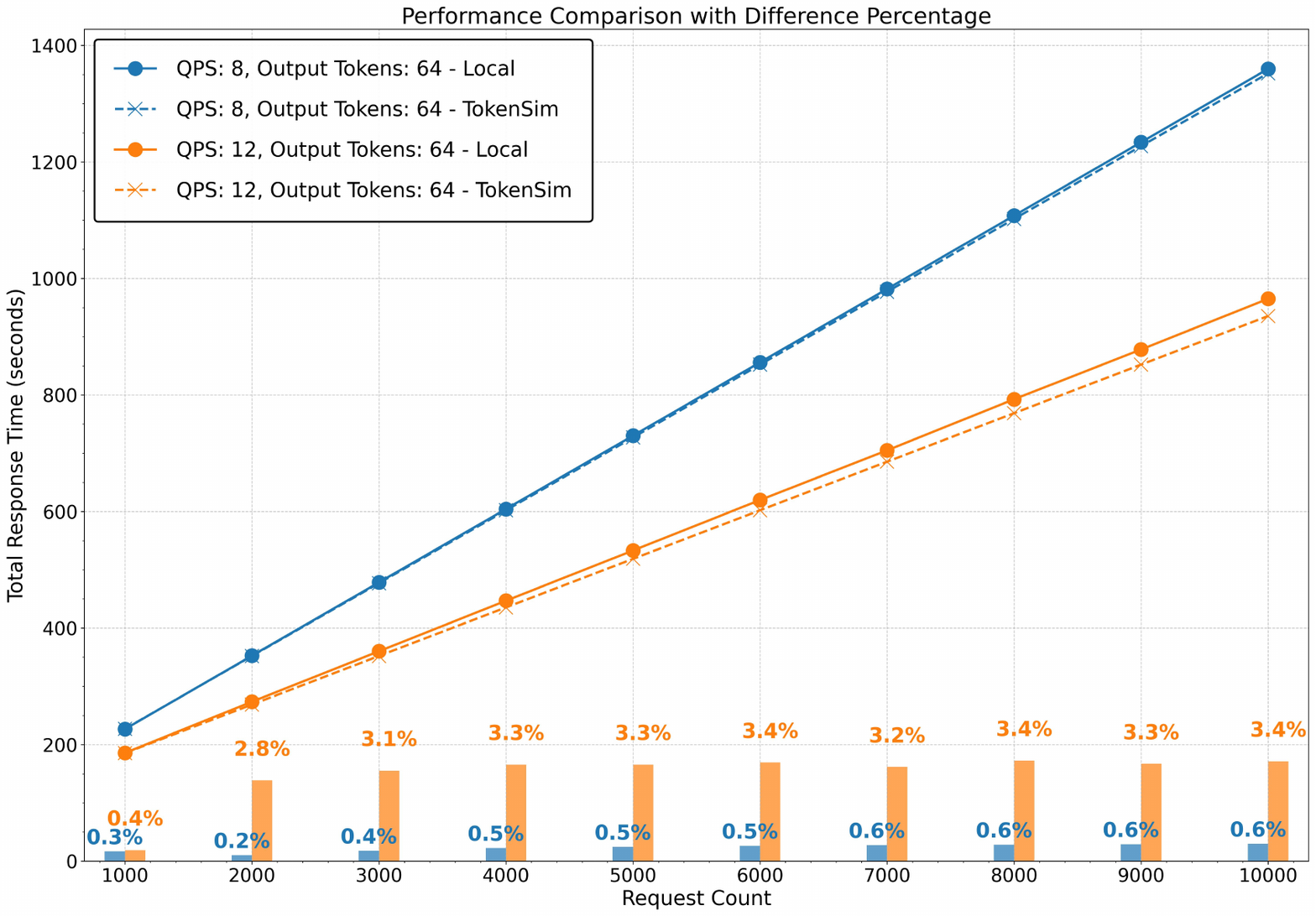}}
    \vspace{-1mm}
    \caption{Runtime comparison between DistServe and \project. The experiment is conducted on 2 A100 GPUs, with each request comprising 64 input tokens and 64 output tokens, representing a common real-world setting. We test the actual communication bandwidth and use it for accurate simulation.}
    \label{fig:distserve_bench}
    % \vspace{-1mm}
\end{figure}

As shown in Figure~\ref{fig:distserve_bench}, \project demonstrates high accuracy in simulating the disaggregated prefill and decoding phases. We attribute the observed discrepancies to two primary sources: 
First, the transmission of KV-Cache over the bus is subject to unavoidable fluctuations, especially under a large volume of requests, which can accumulate and introduce certain errors. 
Second, DistServe employs SwiftTransformer as its underlying runtime framework, which we did not incorporate in our simulation. This difference is an inevitable source of error. 
Additionally, our experiments revealed that the error introduced by SwiftTransformer is more pronounced when the number of requests is relatively low.

Prefilling and Decoding separation technology is widely used in current inference deployments, offering significant performance improvements. The simulation of PD separation by \project highlights its robust performance.

\section{Framework Optimization Analysis with \project}

% In the context of LLM inference, it is crucial to recognize that while many contemporary inference frameworks, such as vLLM\cite{vllm}, provide recommended configurations, performance can significantly fluctuate based on varying workloads and hardware environments. Consequently, optimization strategies, including scheduling methods, batch size modifications, and the disaggregation of prefill and decode processes, may not always enhance performance and could possibly degrade it under certain conditions.

% For instance, tensor parallelism demonstrates superior performance compared to pipeline parallelism on computing setups with high interconnect bandwidth like the HGX platform (with 8x H100 GPUs connected with NVLink). However, this advantage diminishes when applied to systems with PCIe-connected GPUs or across distributed nodes, where inter- and intra-node communication speeds become bottlenecks.

% This section delves into the effects of various optimization strategies on LLM inference, providing a comprehensive analysis of how these strategies interact with different hardware configurations and workload characteristics. By systematically evaluating these factors, we aim to offer insights into achieving optimal performance in diverse computational environments.

In LLM inference, it is crucial to recognize that while frameworks like vLLM \cite{vllm} offer recommended configurations, performance can vary significantly with different workloads and hardware environments. Consequently, optimization strategies such as scheduling methods, batch size adjustments, and the disaggregation of prefill and decode processes may not always enhance performance and could even degrade it under certain conditions.

This section investigates the impact of various optimization strategies on LLM inference, offering a thorough analysis of their interactions with different hardware configurations and workload characteristics. 
Through systematic evaluation, we aim to provide insights into achieving optimal performance across diverse computational environments. Our comprehensive experiments lead to six key findings, which are summarized below.

\subsection{Continuous Batching}\label{cont_batching}

\begin{figure}[htbp]
    \vspace{-2mm}
    \centerline{\includegraphics[width=1.1\linewidth]{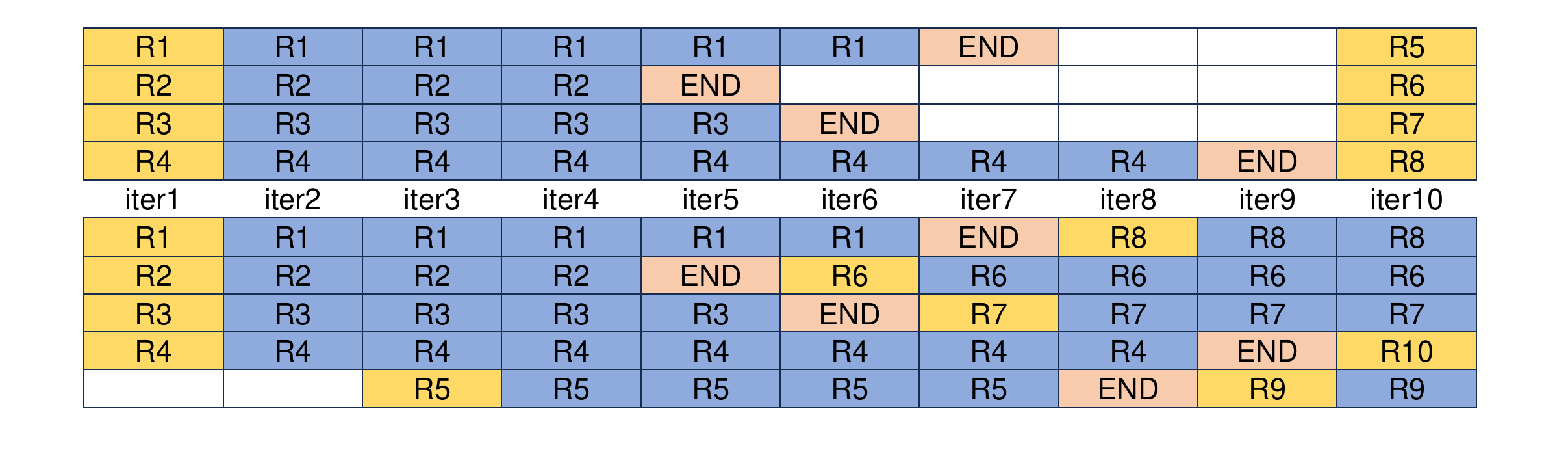}}
    \vspace{-1mm}
    \caption{Comparing iterations of static batching (above) and continuous batching (below). Yellow blocks denote the prefill stage and blue blocks denote the decode stage. White blocks are bubbles.}
    \label{fig:cont_pic}
    % \vspace{-1mm}
\end{figure}

% As depicted in figure \ref{fig:cont_pic}, static batching requires shorter requests to wait for longer ones to complete, resulting in inefficiencies. In contrast, continuous batching allows new requests to be added during the processing of a batch. This flexibility is absent in static batching, where new requests must wait for the current batch to finish, even if the batch size is suboptimal, leading to underutilization of GPU resources. These differences underscore the distinct performance characteristics of the two batching methods and highlight the necessity of simulating continuous batching to accurately predict LLM serving behavior.

Figure \ref{fig:cont_pic} shows an example of the execution when the inference servers takes as input four requests. In static batching, shorter requests have to wait for longer ones to finish, causing bubbles in the system. In contrast, continuous batching allows new requests to be added during batch processing. Thus GPU resources are not wasted in the bubbles. These differences lead to distinct performance characteristics of the two methods and highlight the necessity of simulating continuous batching to accurately predict LLM inference system behavior.

% Figure \ref{fig:static} presents a comparative analysis of static and continuous batching, incorporating PagedAttention mechanisms. The experiment utilized an A100 GPU for the LLaMA2-7B model, processing 50,000 random requests from the ShareGPT dataset\cite{sharegpt}. To effectively demonstrate the unique attributes of continuous batching compared to static batching, the maximum number of batched requests for continuous batching was restricted to match that of static batching. In this setup, a batch size of "inf" indicates no imposed limit, allowing scheduler to maximize resource utilization. Performance was assessed using a normalized latency metric, as evaluated by the vLLM framework \cite{vllm} to demonstrate the overall trend of request latency.

\begin{figure}[htbp]
    \vspace{1mm}
    \centerline{\includegraphics[width=0.5\textwidth]{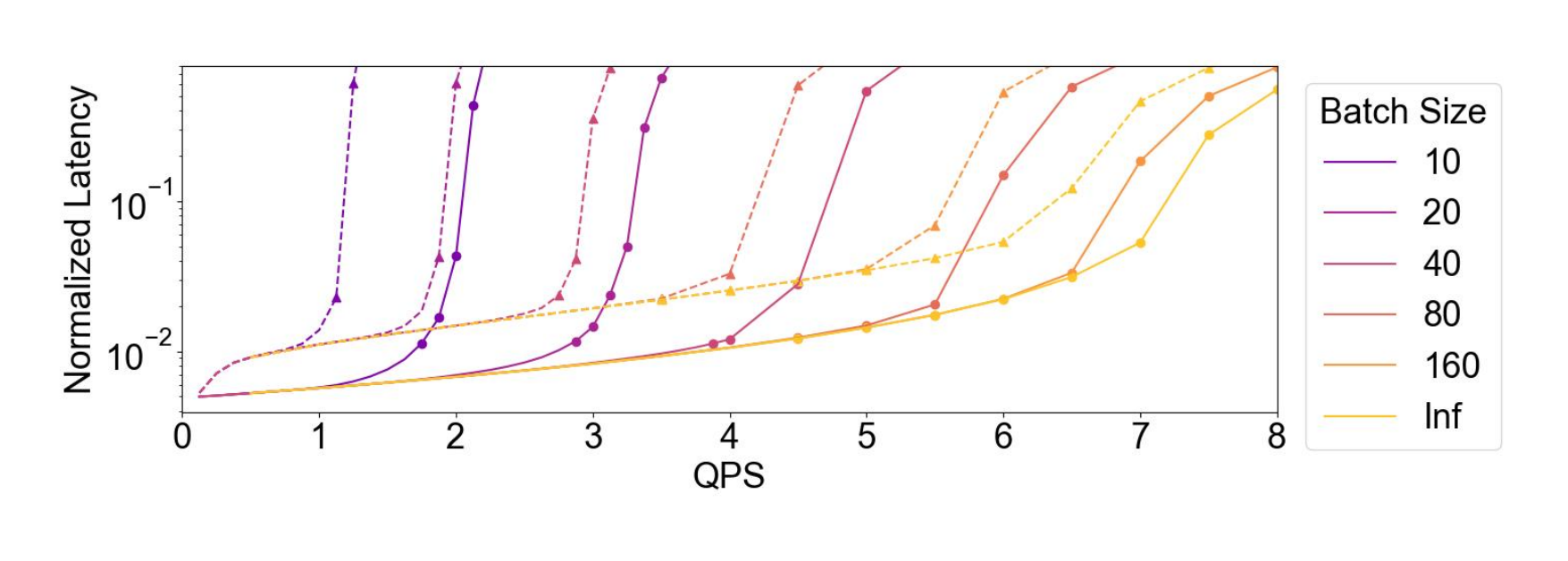}}
    \vspace{-1mm}
    \caption{Normalized latency graph for static batching and continuous batching with limited batch sizes. Dashed lines are static batching and solid lines are continuous batching. "Inf" stands for no batch size limit.}
    \label{fig:static}
    \vspace{-4mm}
\end{figure}

Figure \ref{fig:static} compares the system latency percentile with static and continuous batching. Using an A100 GPU for the LLaMA2-7B model, the experiment processed 50,000 random requests from the ShareGPT dataset. To demonstrate continuous batching's unique attributes, the maximum number of batched requests is restricted to match static batching. A batch size of ``inf'' indicates no limit, allowing the scheduler to maximize resource utilization. Performance is assessed using a normalized latency metric, as evaluated by the vLLM framework \cite{vllm}.

% ===== last finished here

% As depicted in figure \ref{fig:static}, static batching and continuous batching exhibit distinct base latency trends. As the request rate increases, the latency associated with continuous batching rises at a significantly slower and more consistent rate compared to static batching. This observation underscores the critical importance of supporting the simulation of novel batching methods, as continuous batching demonstrates superior scalability and efficiency under increasing load conditions.

As shown in Figure \ref{fig:static}, static and continuous batching exhibit distinct latency trends. As the request rate increases, continuous batching's latency rises at a slower and more consistent rate compared to static batching. 
% This underscores the importance of simulating novel batching methods, as continuous batching demonstrates superior scalability and efficiency under increasing load conditions.
This finding directly supports \textbf{Finding 1}, highlighting that continuous batching significantly reduces latency and improves scalability, especially under increasing load conditions.

\begin{findingbox}
\textbf{Finding 1:} Continuous batching significantly reduces latency and improves scalability compared to static batching, especially under increasing load conditions.
\end{findingbox}

\subsection{Input Batching}\label{ratio}

% Previous research \cite{genz} indicates that increasing batch size can improve throughput with minimal impact on latency, and it suggests that these findings are applicable to continuous batching as well. This implies that maximizing GPU memory usage for KV cache is advantageous in modern frameworks. However, given the significant differences between static and continuous batching, our study seeks to demonstrate that maximizing GPU utilization is not universally optimal across all scenarios.

Previous research \cite{genz} indicates that increasing batch size can improve throughput with minimal latency impact. This implies that maximizing GPU memory usage for KV cache is better for end-to-end performance. 
However, due to differences between static and continuous batching, our study shows that maximizing GPU utilization is not universally optimal.

% For example, when longer requests necessitate additional memory when there's no more free blocks, they preempt newly arrived requests. This process incurs communication overhead due to the need to transfer KV cache, potentially degrading system efficiency. To address this, we propose a method akin to limiting batch size by setting a maximum GPU memory utilization ratio for new requests to be included in the batch. This approach reserves memory space for the potential requirements of older requests, thereby mitigating preemptions.

When longer requests require more memory and no free blocks are available, LLM frameworks often preempt new requests. This involves moving running requests from device memory to host or remote memory.
More importantly, it directly impacts the tail latency because preempted requests usually takes longer than average time to finish.
In conversational scenarios, token generation speed are expected to be fast and evenly distributed. If some tokens take longer, user experience may suffer even if average decode latency meets the Service Level Objective (SLO). When preemptions occur, requests may pause, yet average latency may still comply with the SLO. Therefore, we use the maximum Token Processing Over Time (mTPOT) SLO to highlight the importance of evenly distributed token generation: no interval between tokens should exceed the mTPOT SLO, or the request won't contribute to throughput.

To address this issue, inference frameworks tend to limit batch size by setting a maximum GPU memory utilization ratio for incoming requests, reserving memory for older requests and reducing preemptions. For example, vLLM uses a \textit{gpu\_memory\_utilization} option for performance tunning. 

% In conversational scenarios, the speed of generating new tokens must be both rapid and evenly distributed. If some tokens take significantly longer to generate, the user experience may be disrupted, even if the overall average decode latency remains within acceptable limits. Particularly when preemptions occur, some requests may be temporarily paused, yet the average decode latency may still comply with the Service Level Objective (SLO). Thereby we employ the maximum Token Processing Over Time (mTPOT) SLO to demonstrate how evenly distributed token generation times are crucial: no interval between two tokens should exceed the mTPOT SLO, or the request will not contribute to overall throughput.

% Figure \ref{fig:batch} illustrates the throughput serving requests randomly chosen from shareGPT dataset, where only requests that do not violate the SLO are considered. We set the Time to First Token (TTFT) SLO to 15 seconds and the mTPOT SLO to 0.3 seconds. Figure \ref{fig:batch}(a) depicts the throughput without considering the TTFT SLO, highlighting the latency improvements in the decode stage by restricting new requests. Figure \ref{fig:batch}(b) presents the throughput considering both SLOs, demonstrating the overall effectiveness of this restriction in increasing throughput within SLO.

\begin{figure}[htbp]
    % \vspace{-1mm}
    \centerline{\includegraphics[width=0.5\textwidth]{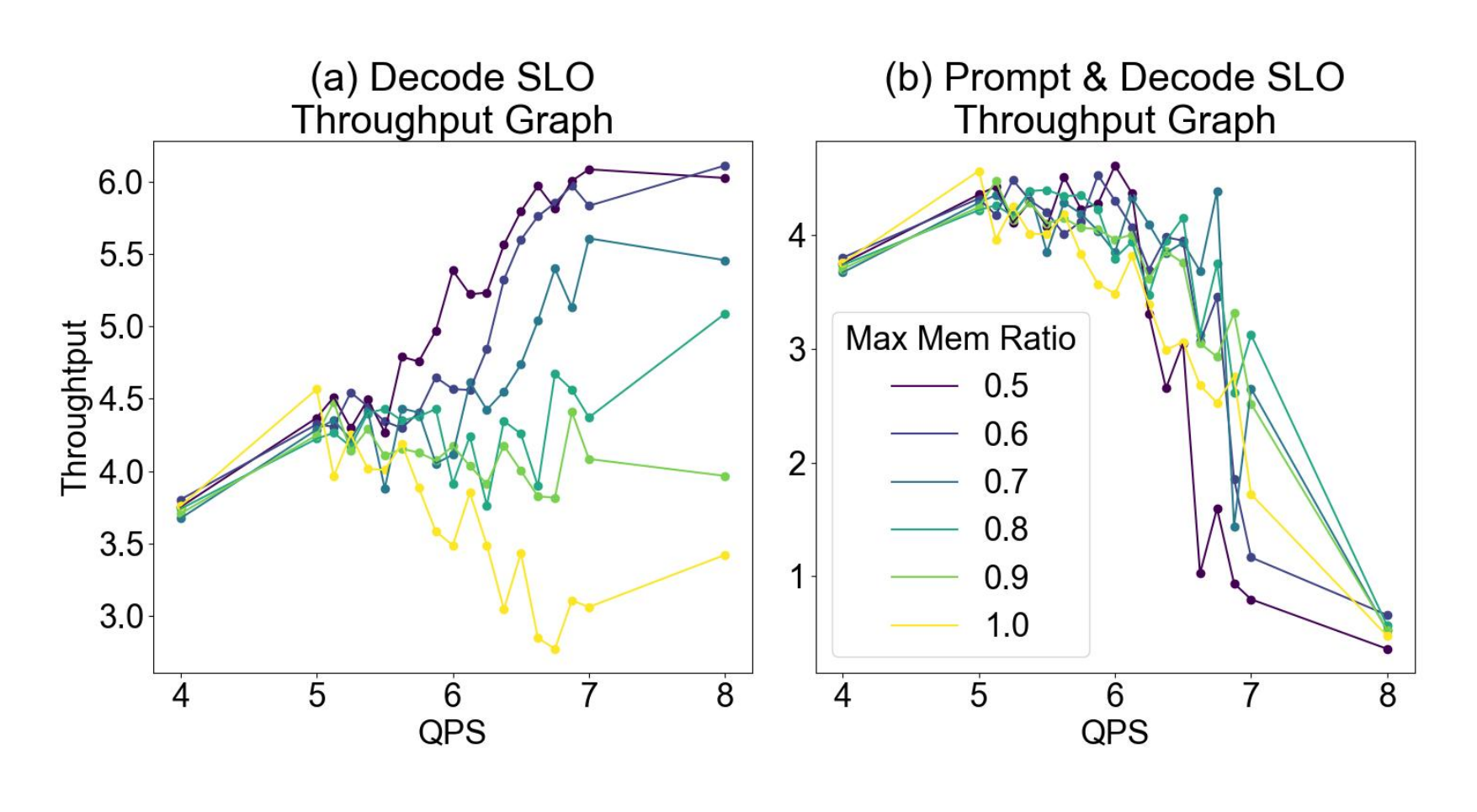}}
    \vspace{1mm}
    \caption{Throughput graphs for restricting GPU memory utilization rate for new request to be scheduled. "Max Mem Ratio" denotes the GPU memory utilization rate, "Decode SLO Throughput" denotes throughput only considering mTPOT SLO, while "Prompt \& Decode SLO Throughput" denotes throughput considering both TTFT and mTPOT SLO.}
    \label{fig:batch}
    \vspace{-2mm}
\end{figure}

Figure \ref{fig:batch} shows the throughput of serving requests from the ShareGPT dataset, considering only those that meet the Service Level Objectives (SLOs). 
We set the Time to First Token (TTFT) SLO to 15 seconds and the mTPOT SLO to 0.3 seconds. Figure \ref{fig:batch}(a) illustrates throughput with only mTPOT SLO, highlighting latency improvements in the decode stage. And figure \ref{fig:batch}(b) presents throughput considering both SLOs, demonstrating the effectiveness of this restriction in increasing throughput within SLOs. 
The results suggest that, under specific conditions, strategically limiting the influx of new requests while reserving memory for previously processed requests—rather than maximizing batch size—can significantly enhance the user experience by optimizing the management of future KV cache.
This supports \textbf{Finding 2}, which states that limiting new requests and reserving memory for older requests can improve tail latency and user experience.

\begin{findingbox}
\textbf{Finding 2:} Limiting the influx of new requests and reserving memory for older requests can enhance user experience by reducing preemptions and improving tail latency, even if it means not maximizing batch size.
\end{findingbox}

% This analysis indicates that, in certain scenarios, limiting the influx of new requests and reserving a portion of memory capacity for older requests to store future KV cache can enhance user experience.

\subsection{Disaggregation Strategies}\label{dis}
\vspace{2mm}
\subsubsection{Device Type Ratio}\label{dis_1}

% Recent studies \cite{splitwise}\cite{distserve} have introduced disaggregated architectures that separate the prefill and decode stages across independent GPUs. However, the impact of varying the ratios of prefill to decode devices remains insufficiently explored. When serving workloads with specific characteristics, these ratios can significantly influence performance outcomes. While some disaggregated architectures \cite{distserve} offer dynamic methods to adjust these ratios during operation, such mechanisms can introduce performance trade-offs.

Recent studies \cite{splitwise, distserve} have introduced disaggregated architectures that separate prefill and decode stages across independent GPUs. However, the impact of varying prefill to decode device ratios is underexplored. These ratios can significantly affect performance outcomes for workloads with specific characteristics. While some frameworks \cite{distserve} offer dynamic methods to adjust these ratios, they can introduce performance trade-offs.

\begin{figure}[htbp]
\vspace{-2mm}
\centerline{\includegraphics[width=1.1\linewidth]{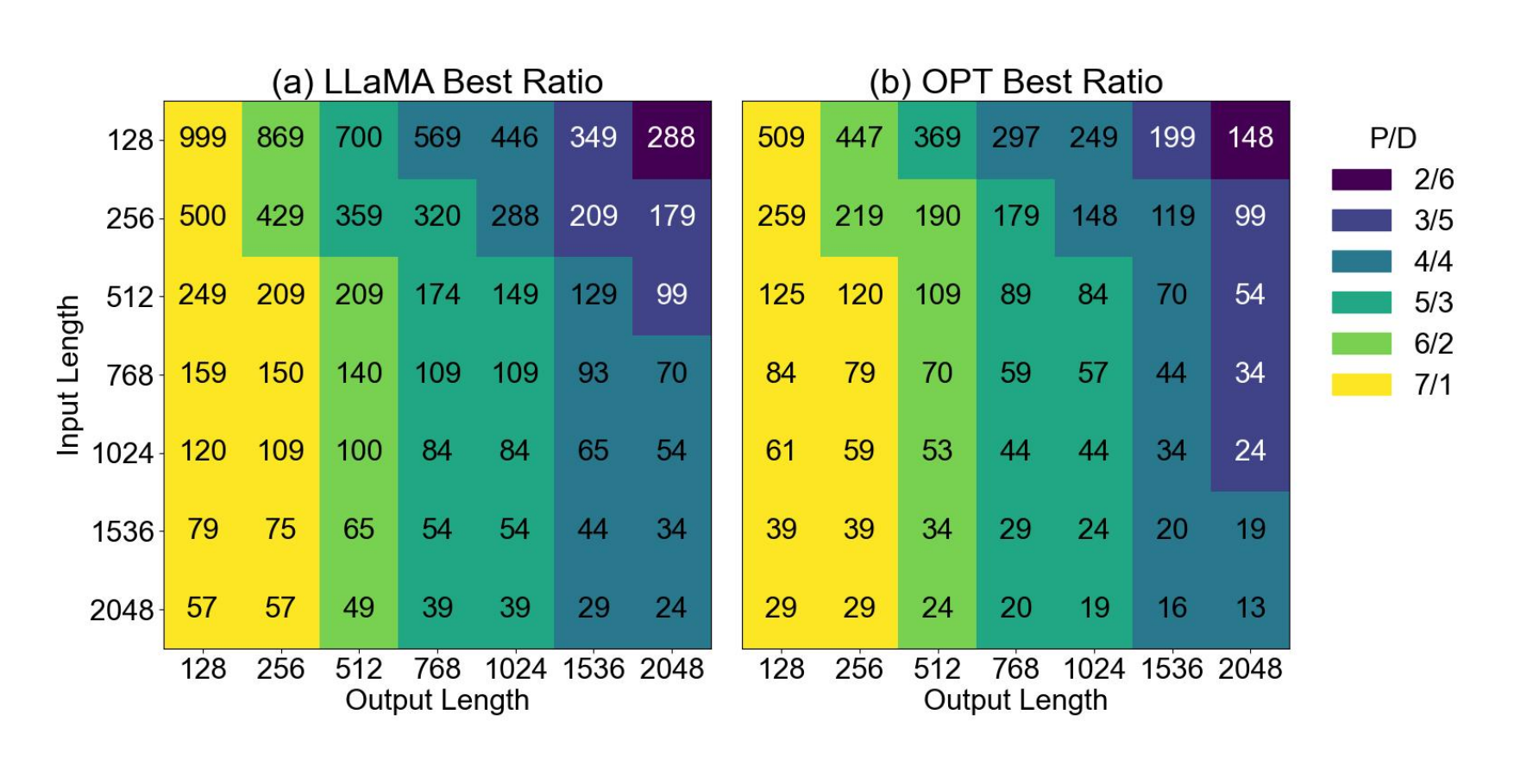}}
 % \vspace{-3mm}
\caption{Best disaggregation type ratio for an 8xA100 node running on different average input and output lengths for LLaMA2-7B and OPT-13B model. P/D denotes prefill and decode devices number. Numbers on the block denotes maximum throughput without violating SLO.}
% \vspace{-2mm}
\label{fig:pd}
\end{figure}

% Figure \ref{fig:pd} illustrates the optimal ratio of prefill and decode devices as different colors on a node with $8\times$ A100 GPUs, across random generated workloads characterized by different average input and output sequence lengths. Focusing on the maximum throughput without violating the serving SLO, the increase of number of prefill nodes seems critical for output lengths, regardless of varying input length. However, when output length increases, smaller input lengths enables more prefill nodes to release for doing decoding.

Figure \ref{fig:pd} shows the optimal prefill to decode device ratio, depicted in different colors, on a node with $8\times$ A100 GPUs across workloads with varying input and output sequence lengths. To maximize throughput without violating the serving SLO, increasing prefill nodes is crucial for longer output lengths, regardless of input length. However, as output length increases, shorter input lengths allow more prefill nodes to be used for decoding.
This analysis supports \textbf{Finding 3}, which highlights that the optimal ratio of prefill to decode devices primarily depends on output lengths, with longer output lengths benefiting more from additional prefill devices.

\begin{findingbox}
\textbf{Finding 3:} The optimal ratio of prefill to decode devices in disaggregated architectures primarily depends on output lengths, with longer output lengths benefiting more from additional prefill devices.
\end{findingbox}

\subsubsection{Efficient Substitutes}\label{dis_2}

% Previous studies \cite{genz}\cite{llmcompass} have identified the decode stage of LLM inference as being memory-bound, heavily reliant on the memory capacity and bandwidth of the GPU. In disaggregated architectures, the prefill and decode stages can be allocated to different hardware components, allowing for optimization based on the specific characteristics of each stage. Given that computational performance is not a critical factor for the decode stage, we conducted a series of tests using various hardware configurations. These configurations included a low compute performance version of the original GPU, a cheaper GPU from a previous generation, and a Processing-In-Memory (PIM) chip, renowned for its high memory capacity and bandwidth.

Previous studies \cite{genz, llmcompass} have shown that the decode stage of LLM inference is memory-bound, relying heavily on the GPU's memory capacity and bandwidth. In disaggregated architectures, prefill and decode stages can be allocated to different hardware components, optimizing based on each stage's characteristics. Since computational performance is not crucial for the decode stage, we tested various hardware configurations, including a low compute performance version of the original GPU, a cheaper GPU from a previous generation, and a Processing-In-Memory (PIM) chip known for its high memory capacity and bandwidth.

\begin{figure}[htbp]
    \vspace{-2mm}
     \centerline{\includegraphics[width=\linewidth]{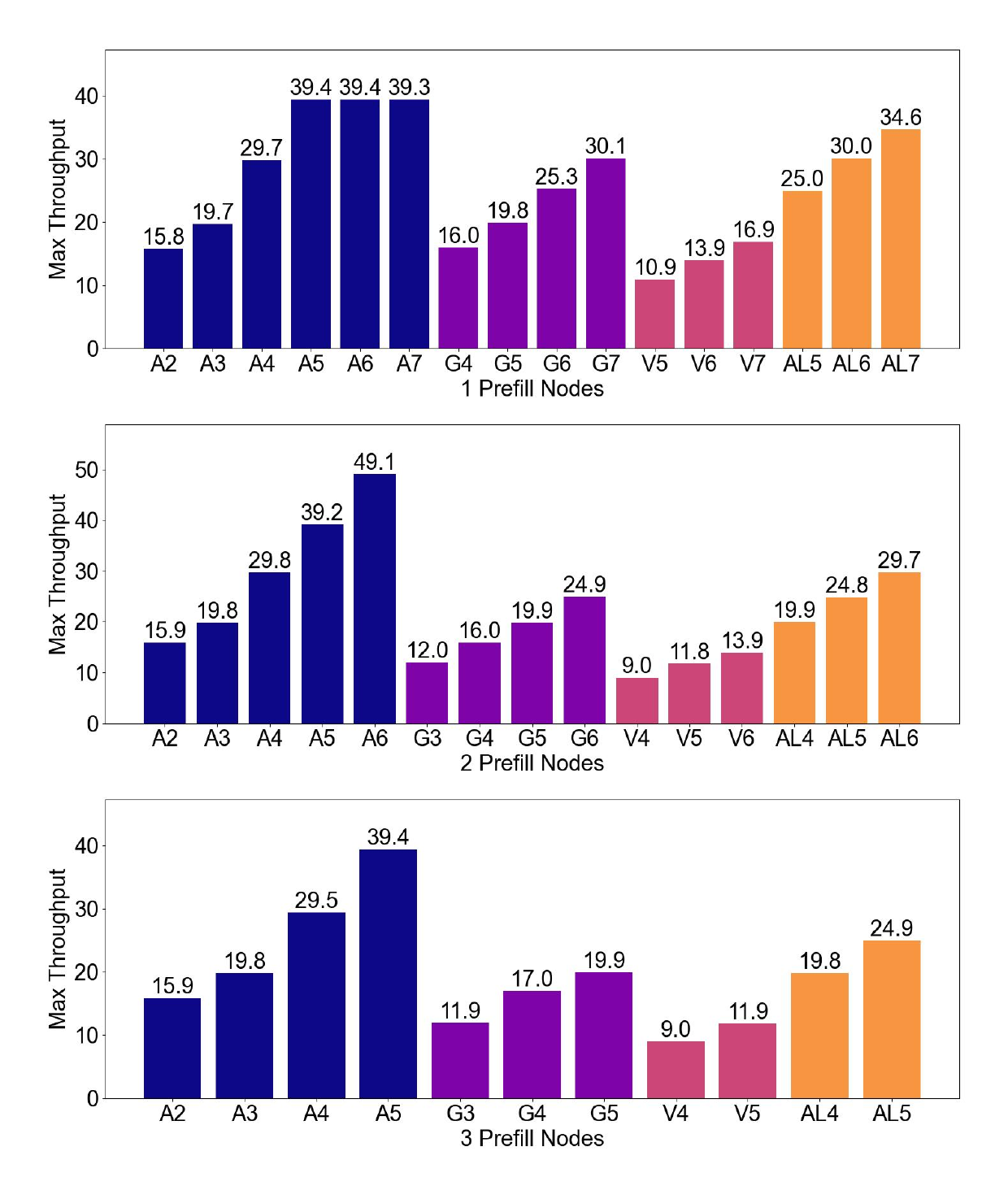}}
     \vspace{-2mm}
    \caption{Disaggregation with Different Hardware: "V" denotes NVIDIA V100, "A" denotes NVIDIA A100, "G" denotes "SK HYNIX GDDR6-Aim", "AL" denotes A100 with $1/4$ peak FLOPS. The number after the letter is the number of decode devices.}
    \vspace{-5mm}
    \label{fig:pd_change_hard}
\end{figure}

% Figure \ref{fig:pd_change_hard} presents the throughput performance of a disaggregated architecture when different hardware devices are used to replace the A100 GPU for the decode stage. Given that most nodes can accommodate a maximum of eight GPUs, we constrained the total number of devices to eight. Three kinds of hardware is tested in this experiment, NVIDIA V100 as a cheaper GPU (about $1/4$ the price of an A100) from the previous generation, SK HYNIX GDDR6-Aim (G6-Aim)\cite{hynix-aim} as a relatively high performance PIM chip (about $1/2$ the price of an A100), and A100 with $1/4$ compute performance.

Figure \ref{fig:pd_change_hard} shows the throughput of a disaggregated architecture when different hardware devices replacing the A100 GPU in the decode stage. Due to the number of PCIe slots available on the server board, we only test configurations with 8 devices. We tested three types of hardware: the NVIDIA V100 as a cheaper GPU (about $1/4$ the price of an A100) from the previous generation, the SK HYNIX GDDR6-Aim (G6-Aim) \cite{hynix-aim} as a high-performance PIM chip (about $1/2$ the price of an A100), and an A100 with $1/4$ compute performance.

% As shown in figure \ref{fig:pd_change_hard}, if the budget only allows for the purchase of 5 A100 GPUs, the optimal choice would be to use 1 A100 as a prefill devices and purchase 7 G6-Aim units. This configuration can save approximately half the cost of an A100 while achieving similar throughput. If 2 A100s are used as prefill devices, the maximum achievable throughput is only $24.7$, which is less than the previous $29.1$. However, if there is a larger budget, using 2 A100s as prefill devices and other A100s as decode devices would always be a better choice. With a smaller budget, such as being able to purchase only 4 A100s, using G6-Aim remains the best option, although the difference is not significant. The performance of the V100 is relatively poor and only holds some value if the budget allows for exactly three or fewer A100s, but the performance difference is not substantial. The performance of a reduced compute performance version of the A100 GPU is suboptimal, indicating that the original A100's computational performance is not excessive for the decode stage.

As shown in Figure \ref{fig:pd_change_hard}, if the budget allows for only $5$ A100 GPUs, the optimal choice is to use $1$ A100 as a prefill device and purchase $7$ G6-Aim units, saving about half the cost of an A100 while achieving similar throughput. If $2$ A100s are used as prefill devices, the maximum throughput is only $24.7$, less than the previous $29.1$. However, with a larger budget, using $2$ A100s as prefill devices and other A100s as decode devices is preferable. With a smaller budget, such as purchasing only $4$ A100s, using G6-Aim remains the best option, though the difference is not significant. The V100's performance is relatively poor, valuable only if the budget allows for exactly $3$ or fewer A100s, but the performance difference is not substantial. The reduced compute performance version of the A100 GPU is suboptimal, indicating that the original A100's computational performance is not very excessive for the decode stage.
In conclusion, PIM can be a cost-effective choice for budget-constrained scenarios but cannot replace the A100 for the decode stage due to the limitation on the total number of slots available within a node.
% This indicates that PIM can be a profitable choice for budget-constrained scenarios but cannot replace A100 for decode stage due to the limitation on the total number of slots available within a node. 

\begin{findingbox}
\textbf{Finding 4:} Processing-In-Memory (PIM) chips can be a cost-effective alternative for the decode stage in disaggregated architectures, especially when budget-constrained, although they cannot fully replace high-performance GPUs like A100 due to slot limitations.
\end{findingbox}

\subsection{Disaggregation Memory Footpoint}\label{pd_mem_trace}

To further investigate the memory requirements of the prefill and decode stages during inference, we conducted an analysis of the GPU memory usage patterns for the LLaMA2-7B model. 
In this experiment, we configured the input length to 128 tokens and the output length to 1024 tokens, using the optimal device allocation and query-per-second parameters identified in Figure~\ref{fig:pd}.
We launched a total of 10,000 requests within a fixed time window of $[5, 65]$ seconds. This time window was chosen because observing the memory footprint only requires analyzing a period of stable operation. 
Additionally, the prefill stage is significantly shorter in duration compared to the decode stage, which allows us to gather sufficient information within this time frame.

\begin{figure}[htbp]
    \vspace{-2mm}
    \centering
    \begin{minipage}[t]{0.49\linewidth}
        \centering
        \includegraphics[width=\linewidth]{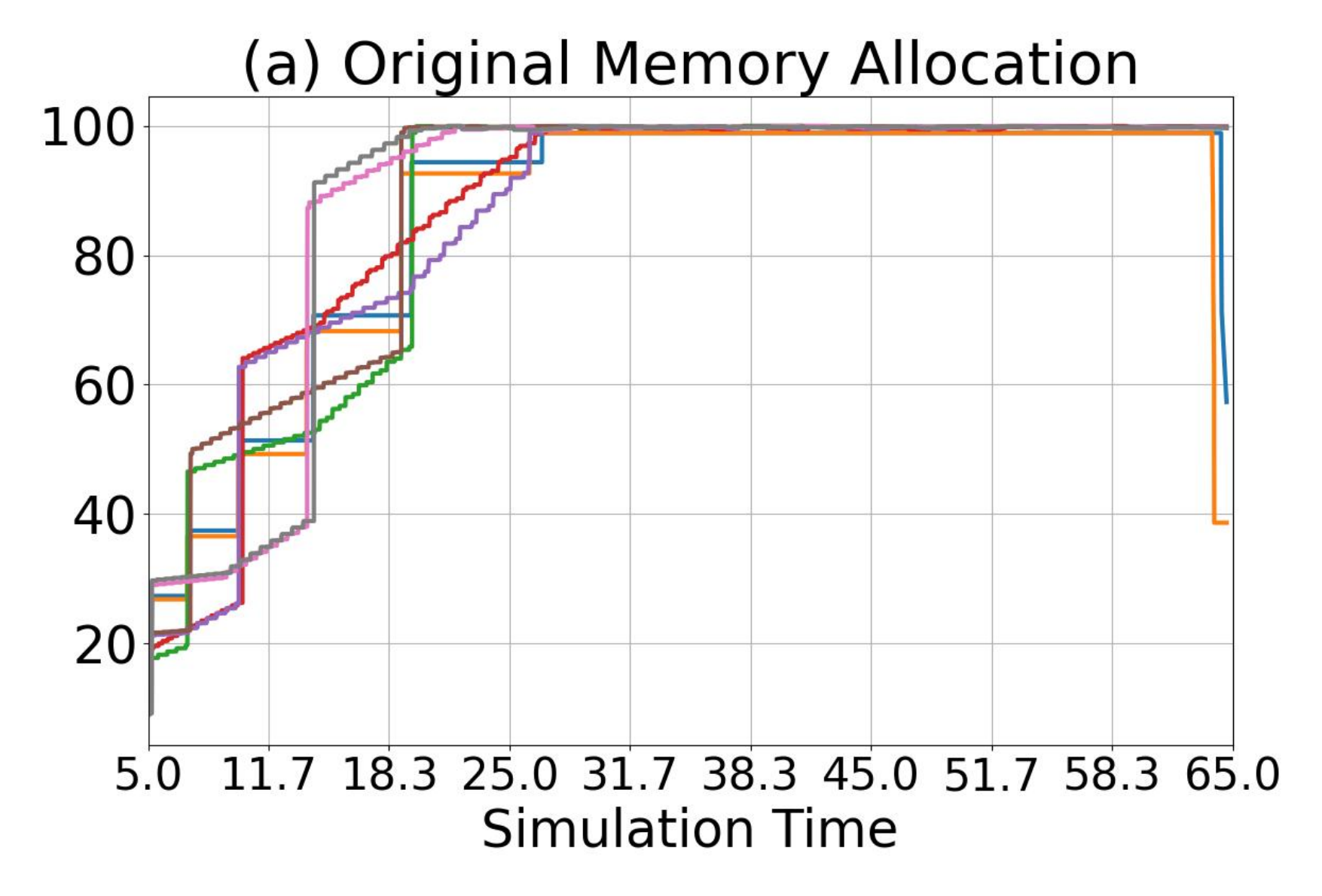}
        
    \end{minipage}
    \hfill
    \begin{minipage}[t]{0.49\linewidth}
        \centering
        \includegraphics[width=\linewidth]{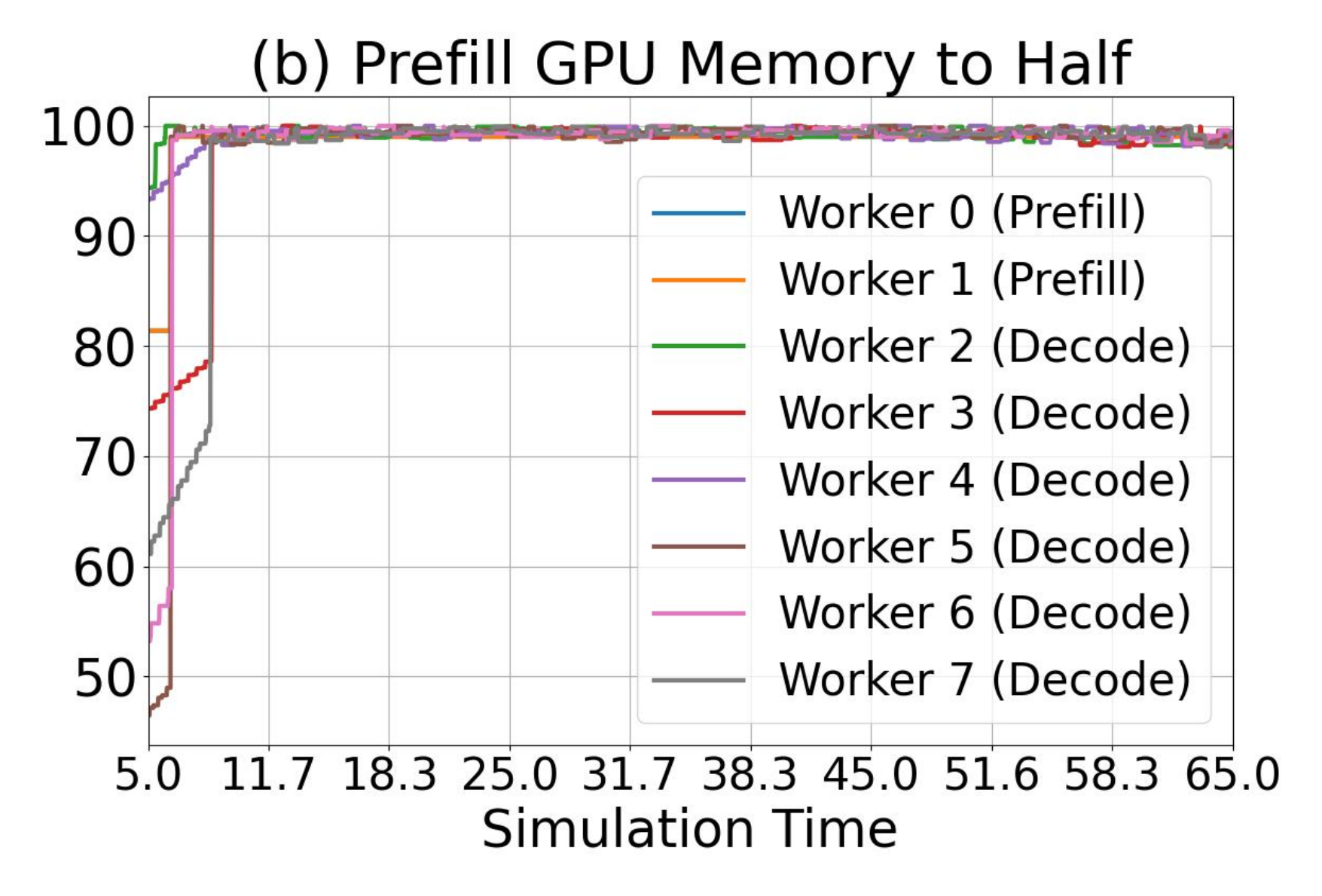}
    \end{minipage}
    \caption{GPU Memory Footprint Heatmaps Over Time for Prefill and Decode Workers in a Disaggregated Architecture. The left heatmap shows the original memory allocation, while the right heatmap demonstrates the effect of reducing the prefill GPU memory to half.}
    \label{fig:pd_mem_trace}
    \vspace{-3mm}
\end{figure}

The memory footprint curves shown in Figure~\ref{fig:pd_mem_trace}(a) reveal that the prefill stage exhibits significantly lower GPU memory usage compared to the decode stage during the PD separation process.
This difference can be attributed to the distinct computational demands of these two stages. Prefill requires substantial initial memory allocation for KV cache computation but maintains lower memory usage thereafter, while decode sustains high memory demand due to its reliance on KV cache for autoregressive token generation. 

Once the prefill stage is completed, the decode devices continue to operate under high memory load, while the prefill devices exhibit significantly lower GPU memory usage.

As highlighted in \textbf{Finding 5}, the prefill stage can be effectively managed with fewer resources, enabling more efficient GPU memory allocation and potentially enhancing system throughput.
We propose that reducing the memory allocation for the prefill GPU may be a viable optimization strategy.

Following this proposal, we reduced the prefill GPU memory to half of its original capacity. 
Figure~\ref{fig:pd_mem_trace}(b) shows reduce prefill GPU memory can achieve a better balance between resource utilization while the throughput remained almost the same.
This finding allows for more efficient allocation of GPU memory and potentially improving overall system throughout.

\begin{findingbox}
\textbf{Finding 5:} The prefill stage exhibits lower memory usage compared to the decode stage. 
Reducing the memory allocation for the prefill GPU may be a viable optimization strategy.
\end{findingbox}

\subsection{Memory Cache}\label{mem}

% Recent studies\cite{memserve}\cite{cachedattention} introduce an innovative approach for serving multi-round conversations, storing the multi-round KV cache in particular storages for further use. However, the specific workloads and hardware configurations can vary significantly among different users, making comprehensive testing under all possible conditions both time-consuming and costly. By incorporating a few additional lines of configurations and codes into \project, it can effectively simulate a simplified version of this mechanism that utilizes a shared memory cache to manage KV cache from contexts of conversations.

Recent studies \cite{memserve, cachedattention} have introduced an innovative approach for serving multi-round conversations by storing the multi-round KV cache in dedicated storage for future use. However, the specific workloads and hardware configurations can vary significantly among users, making comprehensive testing under all conditions time-consuming and costly. By adding a few lines of configuration and code to \project, we can effectively simulate this mechanism that uses a shared memory cache to manage KV cache from conversational contexts.

\begin{figure}[htbp]
% \vspace{-2mm}
    \centerline{\includegraphics[width=\linewidth]{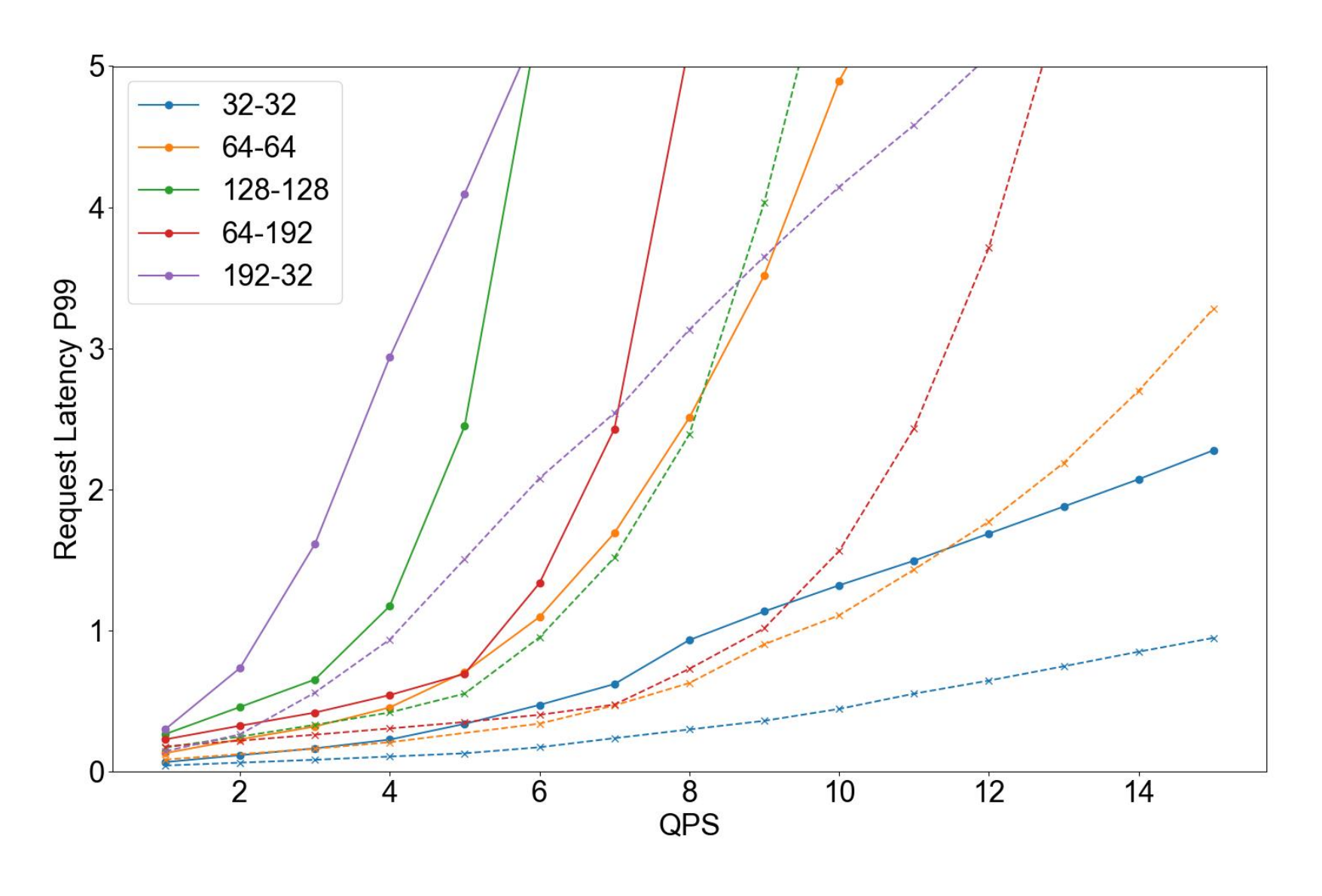}}
    % \vspace{-3mm}
    \caption{Request Latency P99 for memory cache enabled and disabled with different input and output lengths. Legends shown as "x-y" denotes input length is $x$ and output length is $y$. Dashed lines indicates memory cache enabled and solid lines indicates disabled.}
    % \vspace{-1mm}
    \label{fig:mem_pool}
\end{figure}

% Figure \ref{fig:mem_pool} illustrates the P99 request latency across varying input and output lengths, with and without memory cache from multi-round conversations. The conversation lengths tested were generated with a mean length as depicted, following a Poisson distribution. To mimic a realistic chatbot serving scenario, half of the requests were single-round, while the remainder involved two to seven rounds. The latency for retrieving KV cache from the memory pool was set at $800$ nanoseconds per block, as referenced in \cite{memserve}.

Figure \ref{fig:mem_pool} shows the P99 request latency across varying input and output sequence lengths, both with and without the memory cache from multi-round conversations. The conversation lengths are generated with a mean length following a Poisson distribution. To mimic a realistic chatbot scenario, half of the requests are single-round, while the other half involves two to seven rounds. The latency for retrieving KV cache from the memory cache is set at $800$ nanoseconds per block, as referenced in \cite{memserve}.

% The results demonstrate a similar trend to those reported in the original study, while also providing additional insights. As the request rate increases, the use of a memory cache significantly reduces latency, particularly at average output lengths set to 64, effectively doubling the request rate for the same P99 latency. This suggests that the memory cache is most beneficial for serving smaller output lengths, although its advantages diminish for very small outputs, such as those with lengths less than or equal to 32.

The results reveal a trend similar to those in the original study while providing additional insights. As the request rate increases, using a memory cache significantly reduces latency, especially at average output lengths of $64$, effectively doubling the request rate for the same P99 latency. 
This supports \textbf{Finding 6}, which states that memory caching optimizations are most effective for short output lengths in multi-round conversations, significantly reducing latency for outputs around 64 tokens, but offer diminishing returns for very short outputs, such as those with lengths less than or equal to $32$.
Overall, our observation is that memory cache optimizations are most effective for short output scenarios, but always offers advantages over the original version. 

% This indicates memory cache optimizations works best for short output scenarios, but always has a advantage comparing to original version.

\begin{findingbox}
\textbf{Finding 6:} Memory caching optimizations are most effective for short output lengths in multi-round conversations, significantly reducing latency for outputs around 64 tokens, but offer diminishing returns for very short outputs (e.g., $\le$32 tokens).
\end{findingbox}

\section{Platform Characteristics Analysis with \project}
\label{arch}

% In this section, we present a comprehensive analysis of various hardware properties using the \project\ framework. The properties include computing performance, which reflects the ability to execute floating-point calculations; memory bandwidth, which indicates the data transfer speed to and from the GPU's memory; and memory capacity, which determines the amount of data that can be stored in the GPU's memory. We conducted a series of experiments, varying these properties both independently and in combination.

% In this section, we provide an analysis of various hardware properties using the \project\ framework. The examined properties include computing performance, which reflects the ability to execute floating-point calculations; memory bandwidth, indicating the data transfer speed to and from GPU memory; and memory capacity, determining the amount of data that can be stored in GPU memory. We conducted experiments to vary these properties independently and in combination.

In this section, we analyze various hardware properties using the \project\ framework. Our focus is on the following key properties:

\begin{itemize}
    \item \textbf{Computing Performance:} This property reflects the ability to execute floating-point calculations.
    
    \item \textbf{Memory Bandwidth:} This indicates the data transfer speed to and from GPU memory, affecting how quickly data can be moved between the GPU and its memory.
    
    \item \textbf{Memory Capacity:} This determines the amount of data that can be stored in GPU memory, impacting the ability to handle large models and contain more KV cache.
\end{itemize}

In our experiments, we varied these properties both independently and in combination to evaluate their impact on performance. 
While some of our findings align with those reported by existing simulators \cite{genz}, \project have yielded different conclusions when system optimizations are taken into consdieration. Due to space constraints, we present one of these findings in disaggregated settings, with plans for further exploration.

% , but with further optimizations in the framework, diverse workloads, and a broader range of hardware configurations, we have derived several novel insights. Due to space constraints, we present one of these findings in figure \ref{fig:arch_param}, with plans to explore additional results in future works.

\begin{figure}[htbp]
    % \vspace{-1.5mm}
    \centerline{\includegraphics[width=0.5\textwidth]{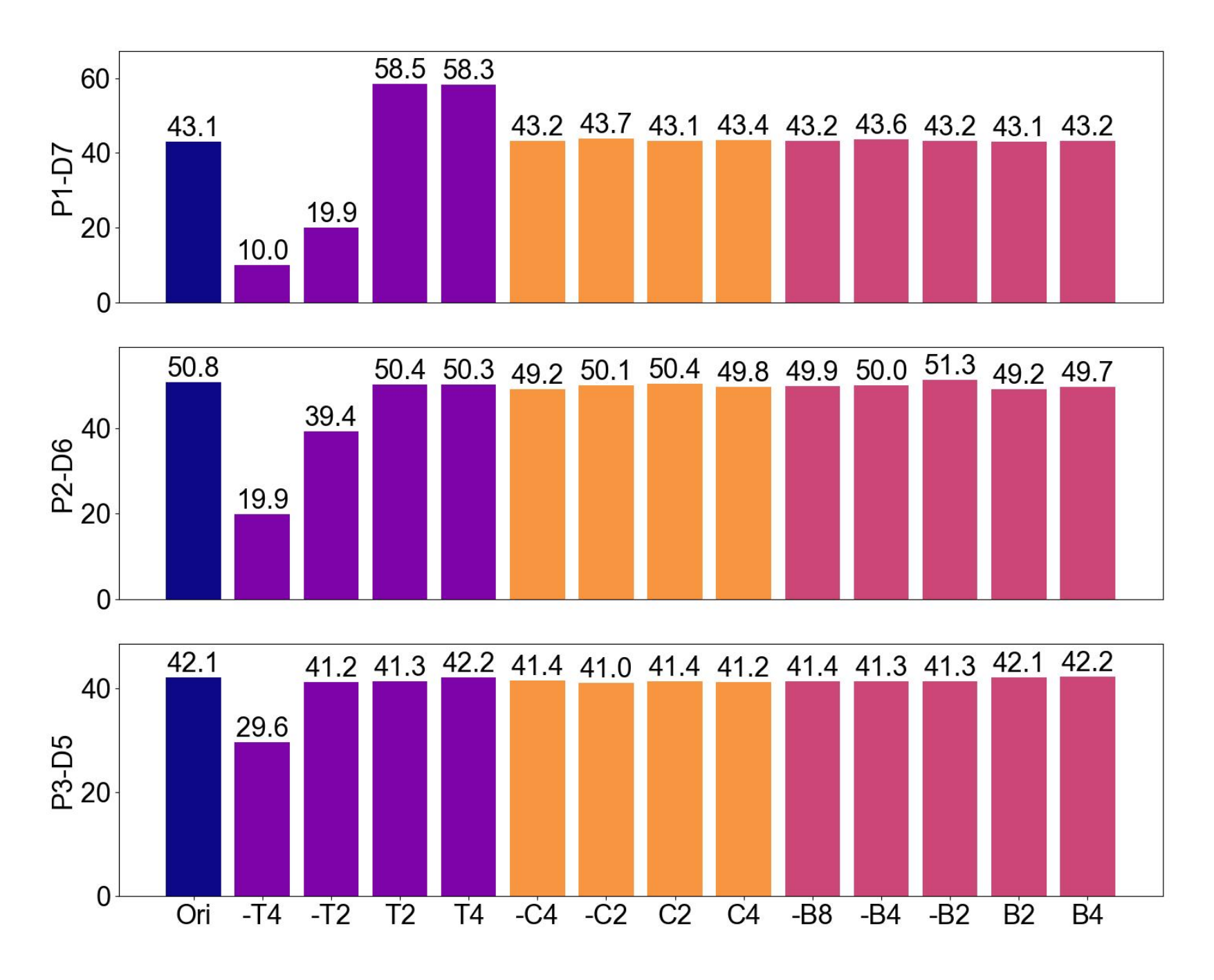}}
    % \vspace{-3mm}
    \caption{Throughput for $1$, $2$, and $3$ prefill devices ("P1-D7" denotes $1$ prefill device with $7$ decode device) with varying hardware parameters: "Ori" denotes original A100; "T" indicates compute performance; "C" and "B" represent capacity and bandwidth, respectively. "C2" doubles capacity while "-C2" halves it.}
    % \vspace{-1mm}
    \label{fig:arch_param}
\end{figure}

% In figure \ref{fig:arch_param}, we evaluated different parameters for the prefill GPU within a disaggregated architecture, processing 50,000 requests from ShareGPT dataset\cite{sharegpt} varying request rates. The figure illustrates the maximum throughput achievable without violating the SLOs. The results indicate that memory capacity ranging from $\times 1/4$ to $\times 4$ and bandwidth ranging from $\times 1/8$ to $\times 4$ of the A100 have minimal impact on performance. In contrast, computational performance significantly influences throughput. However, once the cumulative computational performance reaches $2 \times 312$, it hits the decoding capability limit, beyond which further increases do not enhance throughput.

Figure \ref{fig:arch_param} evaluates different parameters for the prefill GPU within a disaggregated architecture, processing $50,000$ requests from the ShareGPT dataset at varying request rates. 
The figure illustrates the maximum throughput achievable without violating the SLOs. The results show that memory capacity ranging from $1/4$ to $4$ times that of the A100 and bandwidth from $1/8$ to $4$ times have minimal impact on performance, indicating the low bandwidth and capacity requirement of the prefill stage. $1/8$ capacity is untested because it's lower than the model parameter size in fp16. 
In contrast, computational performance significantly influences throughput, but once the cumulative computational performance reaches $2 \times 312$, it hits the decoding capability limit, beyond which further increases do not enhance throughput.
This supports \textbf{Finding 7}, which highlights that the prefill stage benefits more from increased computational performance rather than memory capacity or bandwidth, indicating that the A100 GPU's memory capacity and bandwidth are excessive for prefill tasks.

% This indicates the importance of compute performance for prefill stage and the excessive redundancy of capacity and bandwidth of A100 for doing prefill stage.

\begin{findingbox}
\textbf{Finding 7:} In disaggregated settings, the prefill stage benefits more from increased computational performance rather than memory capacity or bandwidth, indicating that the A100 GPU's memory capacity and bandwidth are excessive for prefill tasks.
\end{findingbox}

\section{Conclusion}

This work introduces \project, a highly extensible framework designed to simulate modern LLM serving systems. \project is designed to work with various hardware setups and system optimizations. Its modular design, along with its support for a wide range of scheduling and memory management techniques, establishes \project as a valuable tool for optimizing LLM inference systems. It shows high accuracy, with an error rate of less than $1\%$ when simulating real-world datasets, further attests to its efficacy. Through the use of \project, we conduct an analysis of current system optimizations, including continuous batching, disaggregated architecture, and memory caching. Furthermore, we explore various hardware configurations, yielding new insights beyond existing work on disaggregated architectures.

% Future work will focus on expanding \project's capabilities and enhancing its scalability to accommodate more model structures like multimodal large language models and more hardware choices including processing-in-memory devices. 
% By making \project fully open source, we aim to foster collaboration and innovation within the research community, encouraging further exploration and development in the optimization of LLM serving systems. 
% This open-source initiative will not only provide researchers with a valuable tool but also contribute to the collective advancement of the field.

% \section*{Acknowledgment}

% \section*{References}

% Please number citations consecutively within brackets \cite{b1}. The 
% sentence punctuation follows the bracket \cite{b2}. Refer simply to the reference 
% number, as in \cite{b3}---do not use ``Ref. \cite{b3}'' or ``reference \cite{b3}'' except at 
% the beginning of a sentence: ``Reference \cite{b3} was the first $\ldots$''

% Number footnotes separately in superscripts. Place the actual footnote at 
% the bottom of the column in which it was cited. Do not put footnotes in the 
% abstract or reference list. Use letters for table footnotes.

% Unless there are six authors or more give all authors' names; do not use 
% ``et al.''. Papers that have not been published, even if they have been 
% submitted for publication, should be cited as ``unpublished'' \cite{b4}. Papers 
% that have been accepted for publication should be cited as ``in press'' \cite{b5}. 
% Capitalize only the first word in a paper title, except for proper nouns and 
% element symbols.

\bibliographystyle{IEEEtran}
\bibliography{main}

\end{document}